\documentclass[preprint,3p,times,authoryear,twocolumn]{elsarticle}
\usepackage{amssymb}
\usepackage{amsmath}
\usepackage{subcaption}
\usepackage{float}
\usepackage{url}
\usepackage{lineno}

\journal{Coastal Engineering}

\begin{document}

\begin{frontmatter}



\title{(Under Review) Hydrodynamics of In-Canopy Flow in Synthetically Generated Coral Reefs Under Oscillatory Wave Motion. \\
\large{[Manuscript has not finished peer-review]}
}


\author[1]{Akshay Patil}
\author[1]{Clara Garc\'{i}a-S\'{a}nchez}
% % AFFILIATION
\affiliation[1]{organisation={3D-Geoinformation Research Group, Faculty of Architecture and Built Environment, Delft University of Technology}, 
addressline={Julianalaan 134},
city={Delft}, 
postcode={2628BL},
country={The Netherlands}
}

%% Abstract
\begin{abstract}
The interaction of oscillatory wave motion with morphologically complex coral reefs showcases a wide range of consequential hydrodynamic responses within the canopy. While a large body of literature has explored the interaction of morphologically simple coral reefs, the in-canopy flow dynamics in complex coral reefs is poorly understood. This study used a synthetically generated coral reef over flat topography with varying reef height and density to understand the in-canopy turbulence dynamics. Using a turbulence-resolving computational framework, we found that most of the turbulent kinetic energy dissipation is confined to a region below the top of the reef and above the Stokes boundary layer. The results also suggest that most of the vertical Reynolds stress peaks within this region positively contribute to the down-gradient momentum flux during the forward phase of the wave cycle. These findings shed light on the physical relationships between in-canopy flow and morphologically complex coral reefs, thereby motivating a further need to explore the hydrodynamics of such flows using a scale-resolving computational framework.
\end{abstract}

\begin{graphicalabstract}
\centering
\includegraphics[scale=0.7,trim={8cm, 1cm, 8cm, 1cm},clip]{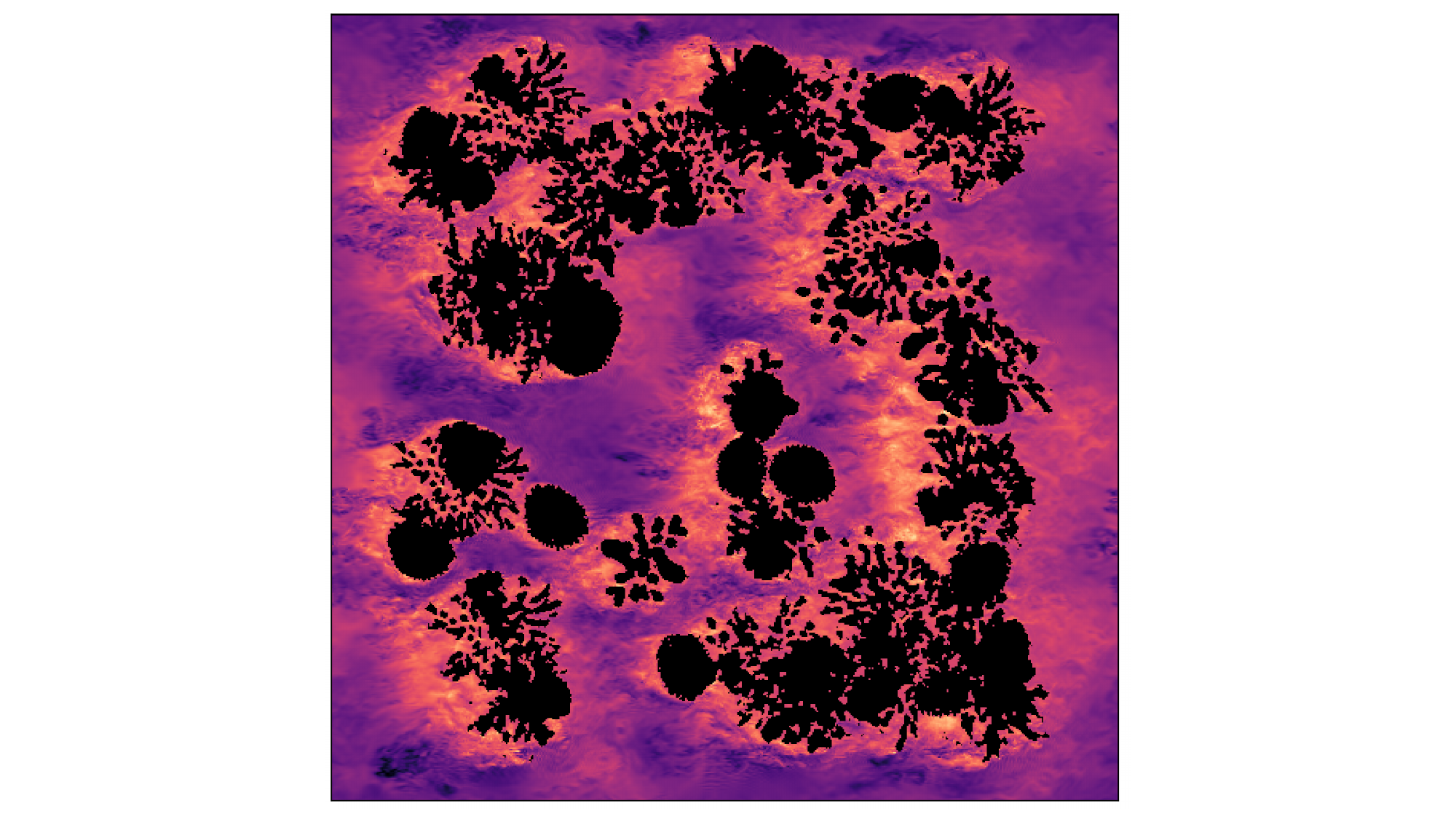}
\end{graphicalabstract}

\begin{highlights}
\item Small-scale turbulent processes are confined to the canopy region where the peak dissipation of the turbulent kinetic energy is observed.
\item The in-canopy vertical Reynolds stress is observed to have peak down-gradient transport during the forward stroke of the wave cycle and vice-versa, respectively. 
\item For synthetically generated coral geometries, the hydrodynamic response is observed to be dependent on the frontal area density in agreement with previous studies.
\end{highlights}

%% Keywords

\end{frontmatter}



\section{Introduction}

Coral reefs exhibit a vital symbiotic relationship with the oceanic environment and its diverse aquatic and biological components, a connection that significantly underpins their overall health and vitality \citep{LoweFalter2015}.  Despite their critical role in the coastal oceanic system, global coral reef health has declined over the past few decades due to increased ocean acidification and warming \citep{Caldeira2003}, even though some coral species have the potential to genetically adapt to the heat stress \citep{Selmonietal2024}. Consequently, there is a growing scientific interest in understanding the life cycle of corals and the hydrodynamic environment they inhabit.  The turbulent oceanic environment provides corals with essential nutrients and supports their overall health and ecosystem. Notably, a pioneering study by \citet{MunkSargent1948} demonstrated that coral reefs flourish prominently in areas characterized by effective wave energy dissipation, as it enables faster cycling of nutrients required by the corals. Leveraging this observation, \citet{Hearnetal2001} proposed a physical model for the nutrient-uptake rates in coral reefs, thus connecting the mass-transfer rate between the hydraulic environment and the underlying benthic organism (i.e., corals) through the turbulent kinetic energy (TKE) dissipation rate. These connections between the hydrodynamics and the coral reef systems motivated a wide range of studies aimed at understanding this crucial ecosystem \citep{Reidenbach2004,Falteretal2005,Reidenbach2006,Monismith2007,RibesAtkinson2007}. \citet{Jonesetal2008} studied the plume dispersion over fringing coral reefs in O'ahu, Hawaii and observed that the later turbulent dispersion coefficients were enhanced due to the interaction of waves and coral roughness. To that end, \citet{NunesPawlak2008} studied the roughness characteristics in O'ahu, Hawaii, to illustrate the connection between the wavenumber spectra and the type of roughness. In another pioneering study to characterise the mass transfer rate as a function of wave strength and coral morphology, \citet{Reidenbach2006} showed that the in-canopy flow increased for both unidirectional and oscillatory flow, thus further enhancing the mass transfer. 

With a wide range of studies illustrating the connection between the coral roughness, hydrodynamics, and mass transfer within the coral canopy, more recent studies have tried to better quantify the in-canopy flow features as a function of the requisite independent parameters \citep[to list a few]{Suzukietal2019,JacobsenMcFall2022,Ascencioetal2022,vanRooijenetal2022}. The central thesis in some of these studies has aimed at modelling the effect of in-canopy turbulence as a porous medium, affecting the flow over it by tuning the inertial and frictional drag parameters. Such flows have been extensively studied as emergent vegetation to consider the transport of momentum and scalars within and outside the canopy, with some of the most notable studies summarised by \citet{Nepf2011}. To enable such a modelling approach, it is important to understand the turbulence dynamics within the canopy better, as illustrated by pioneering works discussing the flow within such systems \citep{Nepf2000,Ghisalberti2002,Monismith2007,Pujol2013}.

Drawing inspiration from the above-mentioned studies and the need to understand the turbulence dynamics within the coral canopies, this study aims to understand the effect of varying coral height and flow strength on the flow within the coral canopy. To achieve this, we model the flow over synthetically generated coral reefs on flat topography to demonstrate the utility of the computational framework. The paper first introduces the details of the computational model and simulation parameters, along with the convergence history of the flow statistics. This is followed by a detailed discussion of the phase-averaged flow response and the turbulence dynamics. At the end, a summary of our conclusions is presented, and future work directions are briefly discussed.
\section{Methodology}

\subsection{Governing equations and discretisation}

 Assuming small wave-steepness, the flow over the coral(s) can be modelled similarly to that of a sinusoidal pressure gradient driving the flow in shallow waters over benthic boundaries \citep{Nielsen1992,Ozdemiretal2014,GhodkeApte2016}. The governing equations are thus given by

\begin{equation}
    \partial_t u_i + \partial_j \left( u_j u_i \right) = -\frac{1}{\rho_0} \partial_i p + \nu \partial_j \partial_j u_i + U_b \omega \cos (\omega t) \delta_{i1} + \mathcal{F}_\text{IBM}, 
\end{equation}

\noindent subjected to the incompressible continuity equation given by

\begin{equation}
    \partial_i u_i = 0,
\end{equation}

\noindent where $t$ is time, $x_i$ represents the coordinate axis where $i=1,2,3$ correspond to the streamwise, spanwise, and vertical directions, respectively, $u_i$ is the velocity vector, $\rho_0$ is the density of the fluid, $p$ is the pressure, $\nu$ is the kinematic viscosity, $U_b$ is the maximum wave orbital velocity, $\omega \equiv 2\pi/T_w$ is the wave frequency where $T_w$ is the wave period, $\delta_{ij}$ is the Kronecker delta, and $\mathcal{F}_{\text{IBM}}$ is the immersed boundary force. 
This paper uses the summation convention for tensor notation such that any two repeating indices are summed over. 
Using the non-dimensional velocity scale as $u_{i_*} \equiv u_i/U_b$, length scale as $x_{i_*} \equiv x_i/ k_s$ where $k_s$ is the representative maximum height of the coral(s), time scale as $t_* \equiv t \omega$, and pressure scaling as $p_* \equiv p/(\rho_0 U_b^2)$ gives the non-dimensional momentum equations (excluding the immersed boundary force) 

\begin{equation}
    \label{eq:nondimNS}
    \partial_{t_*} u_{i_*} + \frac{U_b}{\omega k_s} \partial_{j_*} (u_{j_*} u_{i_*}) = - \frac{U_b}{\omega k_s} \partial_{i_*} p_* + \frac{\nu}{k_s^2 \omega} \partial_{j_*}\partial_{j_*} u_{i_*} + \cos (t_*) \delta_{i1}.
\end{equation}

\noindent Defining $A = U_b/\omega$ as the wave particle semi-excursion length \citep[equation 1.1.2]{Nielsen1992},  equation \ref{eq:nondimNS} can be written as 

\begin{align}
    \partial_{t_*} u_{i_*} + \frac{A}{k_s} \partial_{j_*} (u_{j_*} u_{i_*}) & = - \frac{A}{k_s} \partial_{i_*} p_* \notag \\ &+ \left( \frac{\nu}{k_s U_b} \right) \left( \frac{A}{k_s} \right) \partial_{j_*}\partial_{j_*} u_{i_*} \notag \\ &+ \cos (t_*) \delta_{i1},
\end{align}

\noindent which can be alternatively written after defining the roughness Reynolds number ($Re_k^b \equiv (U_b k_s)/\nu$) and the relative-roughness ($\Gamma \equiv A/k_s$), which is the inverse of the Keulegan-Carpenter number \citep{Nielsen1992} as

\begin{equation}
    \label{eq:finalNonDimNS}
    \partial_{t_*} u_{i_*} + \Gamma \partial_{j_*} (u_{j_*} u_{i_*}) = \Gamma \left( -\partial_{i_*} p_* + \frac{1}{Re_k^b} \partial_{j_*}\partial_{j_*} u_{i_*} \right) + \cos (t_*) \delta_{i1}.
\end{equation}

\noindent Equation \ref{eq:finalNonDimNS} suggests that flow dynamics in the leading order are affected by $\Gamma$ and $Re_k^b$ for this problem setup.
Consequently, in this study, we focus on understanding how changing the relative roughness and the roughness-Reynolds number affects the flow parameters in a wave boundary layer flow over a synthetically generated collection of corals on flat topography. Moreover, it is clear from this form of the equation that the temporal acceleration balances the oscillatory pressure gradient, while the advection of momentum balances the pressure and diffusion of the momentum scaled by the relative roughness and the Reynolds number. 
The emphasis on the dissipation rate is primarily motivated by the limitations in measuring higher-order moments in experimental and/or in-situ measurements; however, decent estimates for the dissipation rate can be made using as few assumptions as possible, thus providing a link between the nutrient transport and the hydrodynamics as discussed in \citet{Hearnetal2001}. 

The momentum and continuity equations are discretised over a staggered grid using a second-order accurate finite difference in space. 
At the same time, the time integration is done using the fractional-step algorithm and a Runge-Kutta method that is third-order accurate in time \citep{Orlandi2000,MoinVerzicco2016}. 
To introduce the roughness elements (i.e., corals) a volume-penalization immersed boundary method (\textit{VIBM}) as introduced by \citet{Scotti2006} is used such that equation \ref{eq:finalNonDimNS} has an additional body force $\mathcal{F}_{{\footnotesize{IBM}}}$ on the right-hand side. 
Using the signed distance field (SDF), the flow is forced to satisfy the no-slip and no-penetration condition at the interface of the geometry (SDF equals 0).  
With the aid of the SDF, close to the solid-fluid interface, the volume fraction is not set to binary values of 0 or 1 and instead smeared across three grid cells continuously such that the velocity steadily approaches the boundary. 
This results in less severe dispersive numerical errors at the cost of local non-enforcement of the no-normal flow through the solid interface.  This is, however, an acceptable compromise given the simplicity of the immersed boundary method and has been validated extensively to demonstrate its validity for high-Reynolds number flows \citep{Scotti2006,YuanPiomello2015}. The advection term is discretised explicitly, while the viscous term is discretised implicitly in the vertical direction to eliminate the strict time-step limitation. 
The pressure Poisson equation is solved efficiently using a Fast Poisson solver since the streamwise and spanwise directions are homogeneous \citep{McKenneyetal1995}. All the simulations were carried out on the Snellius supercomputer on the Genoa partition containing 96 computing cores on a single node. For cases c1, c2, c5, and c6, 8 computing cores were used; for the rest, two nodes with 192 computing cores were used. 

\begin{figure}[t]
    \centering
    \includegraphics[width=\linewidth,trim={1cm 2cm 1cm 2cm},clip]{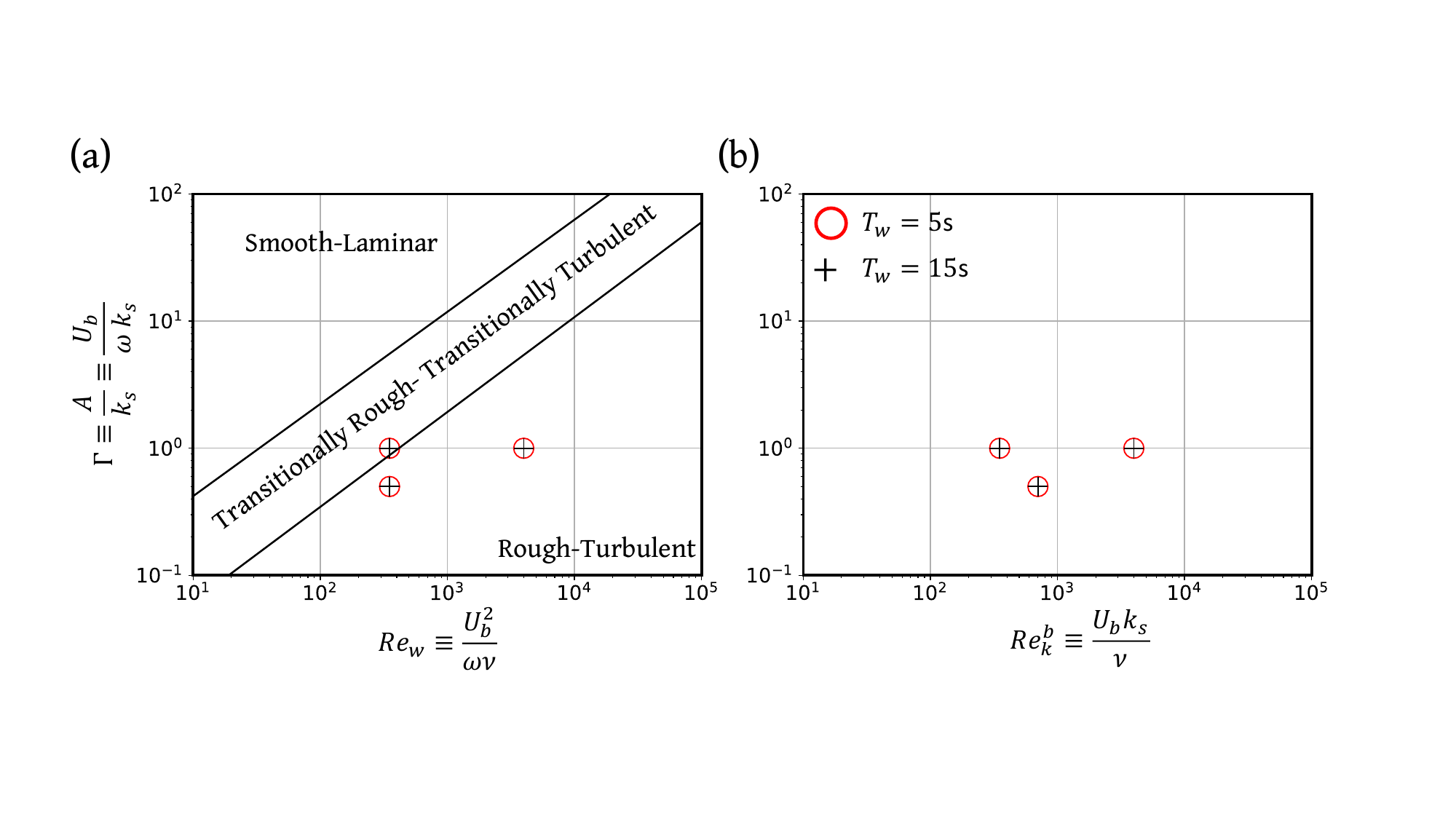}
    \caption{ (a) Simulations carried out in this paper with the wave Reynolds number on the x-axis and the relative roughness on the y-axis. The black solid lines mark the hydraulic and flow classification regimes as discussed in \citet{LacyMacVean2016}. (b) The same data as panel (a) with the roughness Reynolds number on the x-axis. Empty circles mark the cases with a wave period of $T_w = 5$s (i.e., cases c1, c3, and c5), while the plus symbols mark the cases with a wave period of $T_w = 15$s (i.e., cases c2, c4, and c6). }
    \label{fig:figure1}
\end{figure}

To understand the flow response subjected to various synthetic coral collection configurations, we ran 6 simulations as detailed in table \ref{table:table1}. 
Figure \ref{fig:figure1}(a) presents the same data on a phase plot with the x-axis marking the wave Reynolds number (i.e., the strength of the wave) and the y-axis marking the relative roughness ($\Gamma$). 
In this formulation, $\omega$ and $U_b$ appear both on the x- and the y-axis of the parameter space as shown in figure \ref{fig:figure1}. 
As a result, to obtain a consistent non-dimensional timescale (i.e., $\omega$ to shear velocity and viscosity-based time scale), we vary the maximum wave orbital velocity ($U_b$) and the maximum roughness height ($k_s$) to change the non-dimensional parameters. 
One could alternatively vary the viscosity to change the wave Reynolds number; however, since the wave boundary layer thickness also varies with the changes in viscosity, in this case, this option was discarded to have a consistent formulation of the parameters. 
Ultimately, this results in different roughness-Reynolds numbers ($Re_k^b$) for identical inertial wave-Reynolds numbers ($Re_{\omega}^b$) between cases c1, c2, c3, c4, and cases c5, c6, respectively. 
As seen in table \ref{table:table1}, the choice of such a parameter space leads to varying submergence ratios defined as the ratio of the coral roughness to the height of the water column ($k_s/H$), in this case, the height of the water column corresponds to the height of the computational domain. 
As a result, the $x_3$ coordinate of the computational domain is varied such that the top boundary condition does not severely affect the in-canopy flow.

\begin{table*}[t]
    \centering
    \begin{tabular}{l | c | c | c | c | c | c | c | c }
       Case Name & $Re_w \equiv \frac{U_b^2}{\omega \nu}$ & $Re_k^b \equiv \frac{U_b k_s}{\nu}$ & $\Gamma \equiv \frac{A}{k_s}$ & $U_b$ [m/s] & $T_w$ [s] & $k_s$ [m] & $ N_{x_3}$ & $N_{k}$\\ 
       \hline
       c1 & 351 & 351 & 1 & 0.021000 & 5 & 0.0167 & $128$ & $ 40$ \\
       c2 & 351 & 351 & 1 & 0.012125 & 15 & 0.0289 & $128$ & $ 50$ \\ 
       c3 & 3990 & 3990 & 1 & 0.070809 & 5 & 0.0564 & $256$ & $120$ \\
       c4 & 3990 & 3990 & 1 & 0.040882 & 15 & 0.0976 & $256$ & $150$\\
       c5 & 351 & 702 & 0.5 & 0.021000 & 5 & 0.0334 & $128$ & $ 60$\\
       c6 & 351 & 702 & 0.5 & 0.012125 & 15 & 0.0579 & $128$ & $ 60$\\ 
       \hline       
    \end{tabular}
    \caption{Simulations carried out in this paper. For all cases, $512$ grid points are used in the streamwise and spanwise directions while $N_{x_3}$ and $N_k$ correspond to the number of grid points in the vertical and over the maximum roughness height ($k_s$), respectively. Cases c1, c2, c5, and c6 required about 1000 CPU hours each, while cases c3 and c4 required about 92,160 CPU hours each.}
    \label{table:table1}
\end{table*}

\subsection{Coral bed generation}

\begin{figure}[t]
    \begin{subfigure}{.475\linewidth}
        \includegraphics[width=\linewidth]{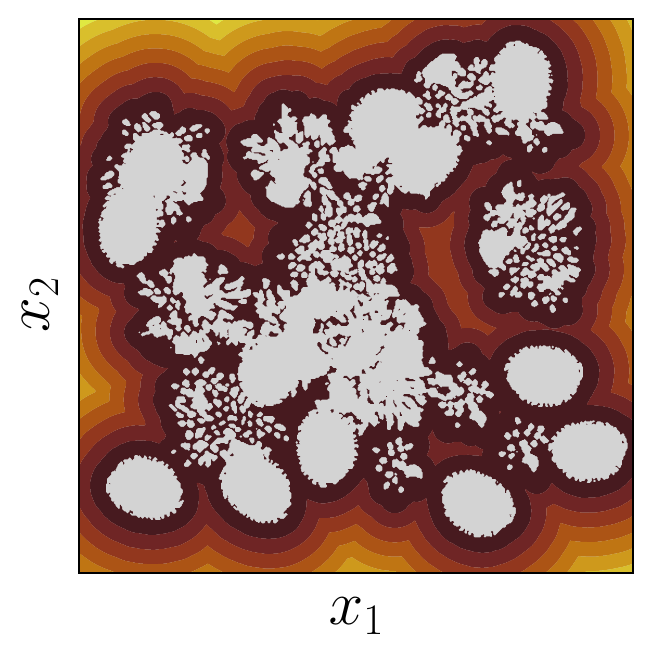}
        \caption{Case c1}
    \end{subfigure}\hfill
    \begin{subfigure}{.475\linewidth}
        \includegraphics[width=\linewidth]{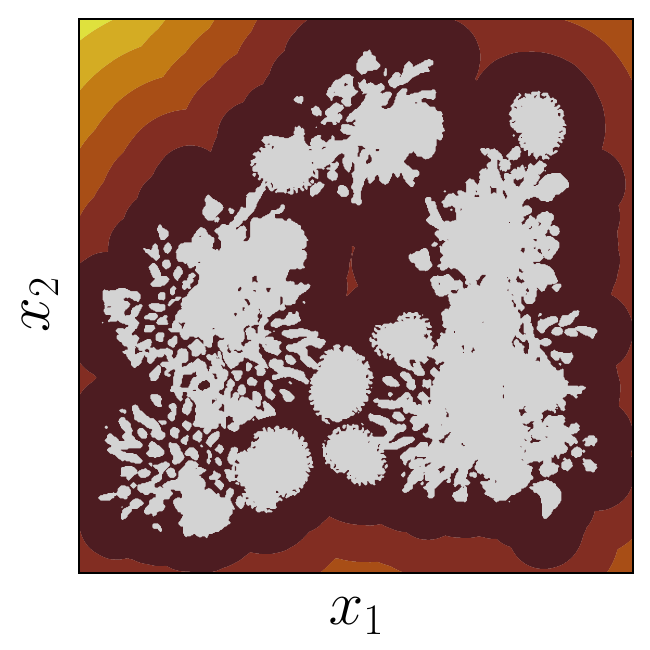}
        \caption{Case c2}
    \end{subfigure}
    \begin{subfigure}{.475\linewidth}
        \includegraphics[width=\linewidth]{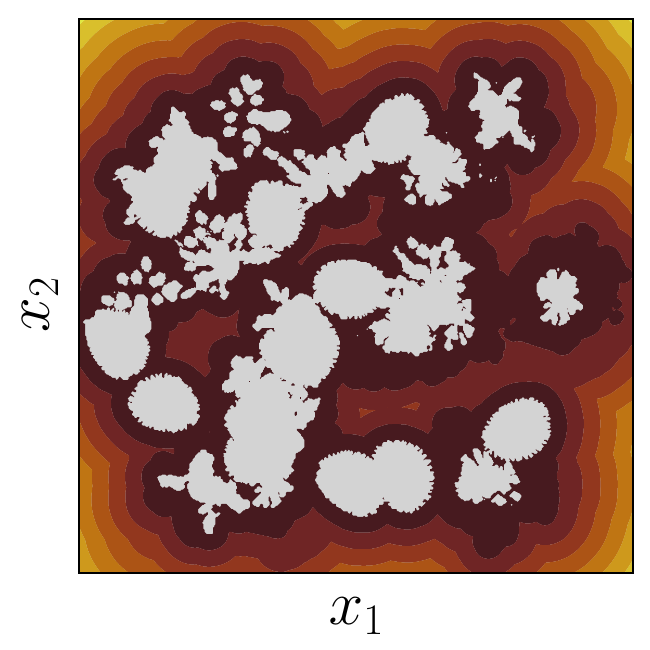}
        \caption{Case c4}
    \end{subfigure}\hfill
    \begin{subfigure}{.475\linewidth}
        \includegraphics[width=\linewidth]{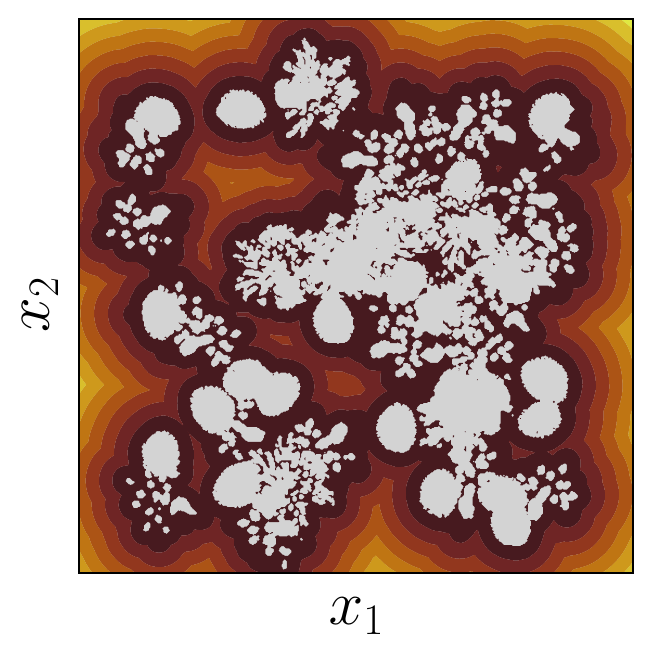}
        \caption{Case c6}
    \end{subfigure}
    \caption{Top view of the stochastically generated corals sliced at $k_s/2$. The light grey colour marks the corals, while the shading marks the positive SDF. }
    \label{fig:figure2}
\end{figure}

The coral bed is generated using a randomly (uniform distribution) sampled collection of three coral geometries, namely \textit{Acropora cervicornis}, \textit{Acropora secale}, and \textit{Goniastrea favulus}. 
The 3D scans for the corals were obtained from the Smithsonian coral scan repository (\url{https://3d.si.edu/corals}) that provides manifold geometries requiring minimal to no geometry repair through the watertight 3D stereolithography (stl) file format. 
The coral geometries are then randomly sampled, translated, and rotated within the defined computational limits to generate a contiguous coral stl. 
The translation, rotation, and coral index choices are uniformly sampled from the prescribed spatial, rotational, and coral index limits, respectively. 
While the individual coral geometries are watertight, the collection of corals generated using the sampling approach is not collectively watertight as it is ambiguous to define watertightness for non-manifold geometries. 
Thus, the collection of corals is wrapped using the wrapwrap (\url{https://github.com/ipadjen/wrapwrap}) application built on the alpha-wrap algorithm developed by \citet{cgalalphawrap}. Choosing the input values for the relative alpha ($\alpha = 1500.0$) and relative offset $(\beta = 2000.0)$, wrapwrap generates a single watertight coral geometry and reduces the total file size, making the SDF computation less memory intensive.  The watertight geometry ensures a non-degenerate SDF using the stl2sdf generator, which uses a local sampling method to generate the SDF. Detailed scaling results for stl2sdf are discussed in \ref{app:appendixB}. Figure \ref{fig:figure2} illustrates the typical stochastically generated corals used for four of the six cases in this paper. Using the geometry translation, the maximum height of the corals can be exactly set as detailed in table \ref{table:table1}. This computational workflow allows the generation of stochastic coral reefs over flat terrains with specified extents and maximum height. The code has been since updated to handle non-flat topography using Perlin noise \citep{perlin1985}; however, in this paper, we will only discuss corals over flat terrains.

\begin{figure}[H]
    \centering
    \includegraphics[width=\linewidth]{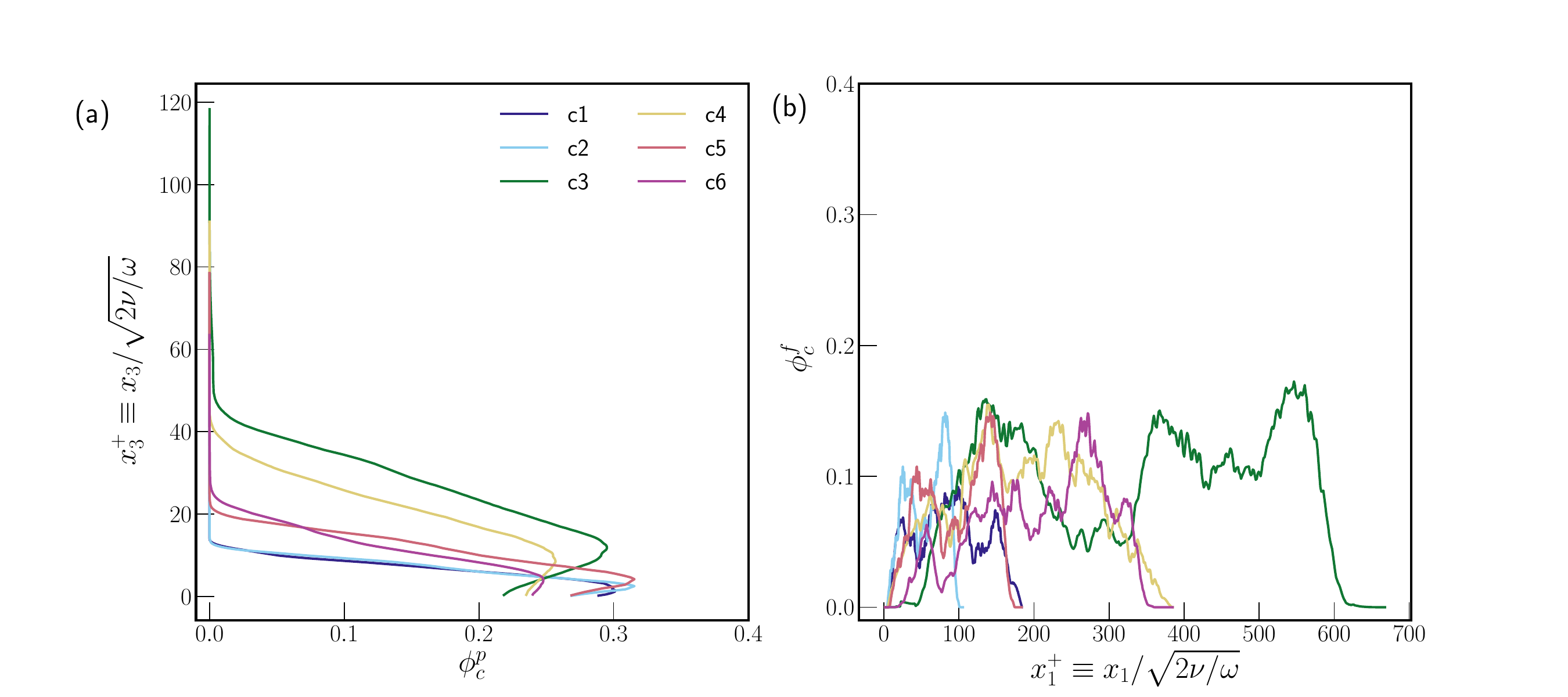}
    \caption{(a) Planform area fraction occupied by the coral (solid) as a function of the vertical coordinate. The two columns in the legend correspond to the varied relative roughness parameter ($\Gamma$) where the first column has $\Gamma = 1.0$ and the second column has $\Gamma = 0.5$ for identical wave Reynolds number. (b) Frontal area fraction occupied by the coral (solid) as a function of the streamwise coordinate. Color scheme for the lines same as panel (a).}
    \label{fig:figure4}
\end{figure}

Figure \ref{fig:figure4}a shows the vertical profile of the planform solid fraction ($\phi_c^p$); for all cases, $\phi_c^p$ peaks close to 0.3. Here, the platform solid fraction is defined as the ratio of the area occupied by the coral to the total available platform area at a given distance above the wall. Figure \ref{fig:figure4}b shows the streamwise profile of the frontal solid fraction ($\phi_c^f$) where, for different cases, the peak value varies from 0.1 to 0.3. In addition to the coordinate-dependent area fractions, the frontal and planform projected areas of the solid fraction (corals in this case) are important indicators. Thus, table \ref{table:table2} presents a detailed summary of the frontal and planform projected areas for the various cases. It is clear from table \ref{table:table2} that despite the control over the height and layout, there is a large variation in the frontal area occupied by the corals where it ranges from about 0.2850 (minimum) to 0.6235 (maximum). As a result, despite the tight control over the coral height, changes in all the requisite parameters of interest that affect the flow dynamics cannot be directly controlled when designing a parametric sweep.

\begin{table}[H]
\centering
    \begin{tabular}{l | c | c  }
       Case Name & $A_c^p/A^p$ & $A_c^f/A^f$ \\
       \hline 
       c1 & 0.4456 & 0.2850 \\
       c2 & 0.5250 & 0.4273 \\
       c3 & 0.4638 & 0.6100 \\
       c4 & 0.4655 & 0.5764 \\
       c5 & 0.5250 & 0.4303 \\
       c6 & 0.4207 & 0.6235 \\
       \hline       
    \end{tabular}
    \caption{Non-dimensional planform and the frontal area occupied by the corals for the cases simulated in this paper.}
    \label{table:table2}
\end{table}

\subsection{Convergence of turbulence statistics and additional parameters}

All the cases are run for a total of 60 wave periods, with the results being stored at every $\frac{T_w}{20}~s$ to capture the turbulence statistics over the entire wave period sufficiently. To understand the convergence of turbulent statistics, we use the volume-averaged estimates of the various flow quantities of interest. For a given flow quantity of interest $f_i$, the planform-average is defined as

\begin{equation}
    \langle f_i \rangle (x_3,t) = \frac{1}{A_f(x_3)} \int_{A_f(x_3)} f_i (x_1,x_2,x_3,t) dA, 
\end{equation}

\noindent where $A_f$ is the planform area occupied by the fluid (i.e., the area occupied by the non-light gray region in figure \ref{fig:figure2}) and the volume-average is the vertical-integral of the planform-averaged quantity given by

\begin{equation}
    \langle f_i \rangle_v (t) = \frac{1}{H} \int_0^H \langle f_i \rangle (x_3',t) dx_3',
\end{equation}

\noindent where $H$ is the height over which the vertical integration is carried out. Additionally, the flow quantity $f_i$ can be decomposed as

\begin{equation}
    \label{eq:velocityDecomposition}
    f_i (x_1,x_2,x_3,t) = \tilde{f}_i (x_1,x_2,x_3,\omega t) + f_i' (x_1,x_2,x_3,t),
\end{equation}

\noindent where the first term on the right-hand side is the phase averaged component of $f_i$ while the second term on the right-hand side is the turbulent component of $f_i$ and the phase average is given by

\begin{align}
    \label{eq:phaseAverageDefinition}
    \widetilde{f}_i(x_1,x_2,x_3,\omega t) & = \frac{1}{N_w} \sum_{j=1}^{N_w} f_i(x_1,x_2,x_3,t+jT_w) \notag \\ & - \overline{f}_i(x_1,x_2,x_3),
\end{align}

\noindent where $\overline{f}_i(x_1,x_2,x_3)$ is the time averaged flow quantity. The phase-averaging component in equation \ref{eq:velocityDecomposition} can be further decomposed using the planform-averaging operator. Thus, equation \ref{eq:velocityDecomposition} can be written as

\begin{align}
    \label{eq:tripleVelocityDecomposition}
    f_i (x_1,x_2,x_3,t) & = \langle \tilde{f}_i \rangle (x_3,\omega t) + \hat{f}_i(x_1,x_2,x_3, \omega t) \notag \\ & + f_i' (x_1,x_2,x_3,t),
\end{align}

\noindent where the second term on the right-hand side of equation \ref{eq:tripleVelocityDecomposition} is the dispersive velocity component defined as 

\begin{equation}
    \label{eq:dispersiveComponent}
    \hat{f}_i(x_1,x_2,x_3, \omega t) = \widetilde{f}_i (x_1,x_2,x_3,\omega t) - \langle \widetilde{f}_i \rangle (x_3, \omega t).
\end{equation}

\noindent In this paper, the triple velocity decomposition as defined in equation \ref{eq:tripleVelocityDecomposition} will be used to understand the phase variability and compute the turbulence statistics. Specifically, the sum of the dispersive and turbulent components will be used to estimate the convergence of the turbulence statistics. This is done in order to understand how the relevant scale of the flow converges as a function of the number of samples (i.e., the number of waves) as the mean flow converges relatively quickly when compared to the fluctuating and dispersive components. Additionally, the phase- and planform-averaged velocity components can be computed once the three-dimensional velocity and pressure fields are obtained. Using the velocity decomposition proposed in equation \ref{eq:tripleVelocityDecomposition}, the sum of the turbulent and the dispersive components can be readily computed as a function of time.

\begin{figure}[t]
    \centering
    \includegraphics[width=\linewidth,trim={4cm 0cm 4cm 0cm},clip]{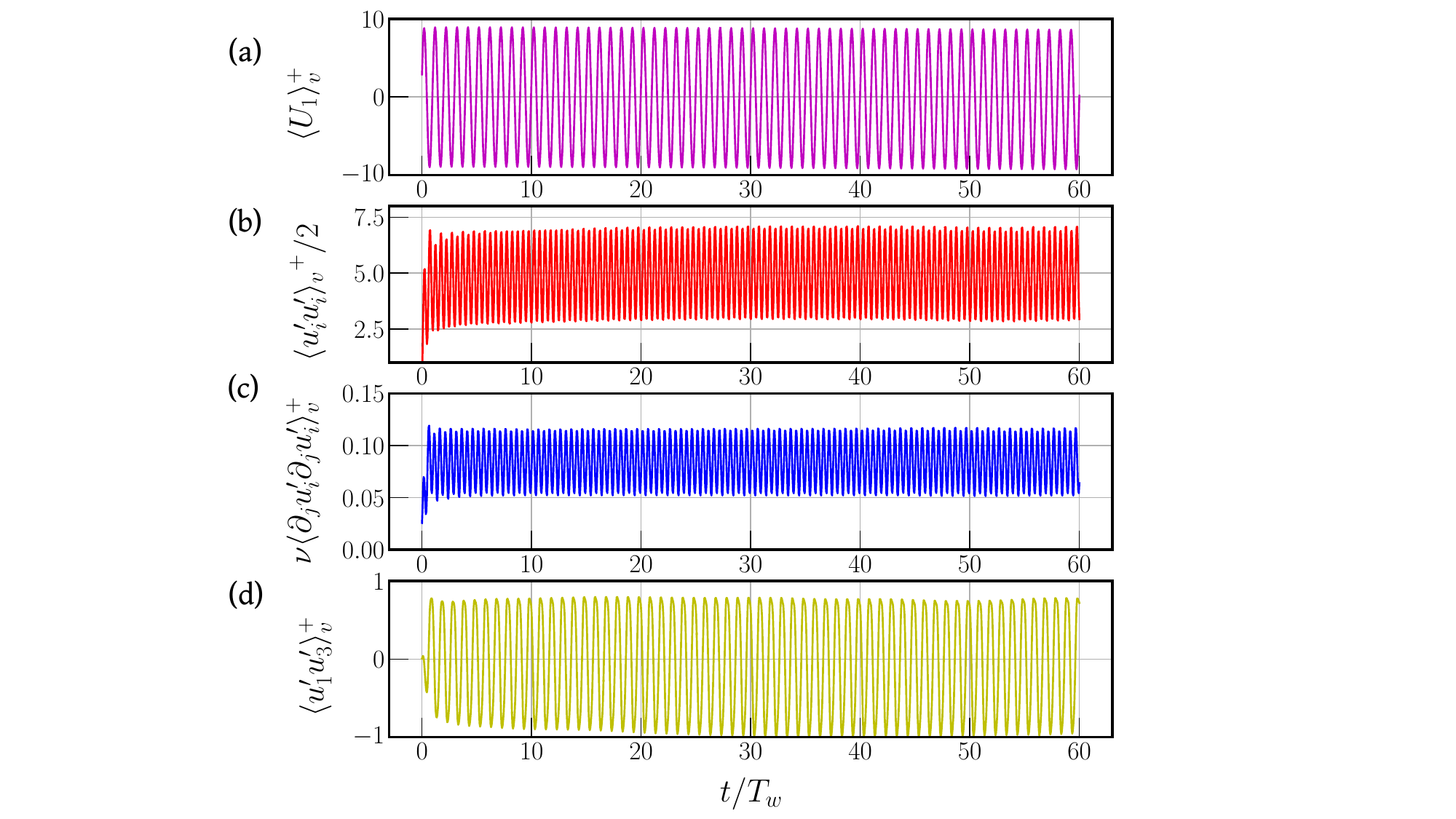}
    \caption{Time evolution of the volume-averaged flow statistics for case c2. All panels are volume-averaged quantities: (a) wave velocity, (b) turbulent kinetic energy (TKE), (c) TKE dissipation rate, and (d) Reynolds stress. The X-axis for all the panels corresponds to the non-dimensional time normalised using the wave period (i.e., the x-axis corresponds to the number of wave periods). Panel (a) is normalised using $u_{\tau}$, panels (b) and (d) are normalised using $u_{\tau}^2$, and panel (c) is normalised using $u_{\tau}^4/(\kappa \nu)$. For case c2, the friction velocity is $u_{\tau} = 0.0024~m/s$ using equation \ref{eq:frictionVelocity}.}
    \label{fig:figure3}
\end{figure}

The time evolution of the various flow quantities of interest for case c2 can be seen in figure \ref{fig:figure3} where the $+$ marker denotes non-dimensionalisation using the maximum friction (i.e., or shear) velocity ($u_{\tau}$), the kinematic viscosity ($\nu$), and the von K\'{a}rm\'{a}n constant ($\kappa$) taken to be $0.41$. The friction velocity for all the cases is defined as 

\begin{equation}
    \label{eq:frictionVelocity}
    u_{\tau} = \sqrt{\text{max} \left[ -\langle u_1' u_3'\rangle_v (t) \right]}.
\end{equation}

\noindent Based on the temporal evolution of the mean flow and the turbulence statistics as shown in figure \ref{fig:figure3}, case c2 converges after 30 wave periods for both the mean flow and the turbulence statistics. The TKE and the TKE dissipation rate converge relatively quickly by around 15 wave periods, however, the correlations between the streamwise and vertical velocity components take relatively longer to converge as seen in figure \ref{fig:figure3} in addition to exhibiting some variations over time. As a result, when presenting the phase averages, the first 30 wave periods for all the turbulence statistics are discarded, and the phase averages are computed for the last 30 wave periods. Similar trends were observed for all the other cases detailed in table \ref{table:table1} as case c2; hence, for the sake of brevity, the convergence history for all the other cases will not be discussed. As detailed in \ref{app:appendixA}, the TKE dissipation rate scales with a pre-factor $\Gamma/Re_k^b$; as a result, it is expected that for identical values of $\Gamma$, the TKE dissipation rate contribution decreases with increasing $Re_b^k$. However, as discussed in \citet{Pomeroy2023}, the area fraction of the corals also plays an important role in flow response above and within the canopy. As a result, the solid fraction, projected frontal, and planform areas are computed for each of the simulation cases carried out in this paper as shown in figure \ref{fig:figure4} and detailed in table \ref{table:table2}. Here, $A_c^p$ is the planform area occupied by the corals, $A^p$ is the total available planform area, $A_c^f$ is the frontal area occupied by the corals, and $A^f$ is the total frontal area available. 

\section{Results and Discussions}

\subsection{Phase averaged velocity}

\begin{figure}[t]
    \centering
    \includegraphics[width=\linewidth]{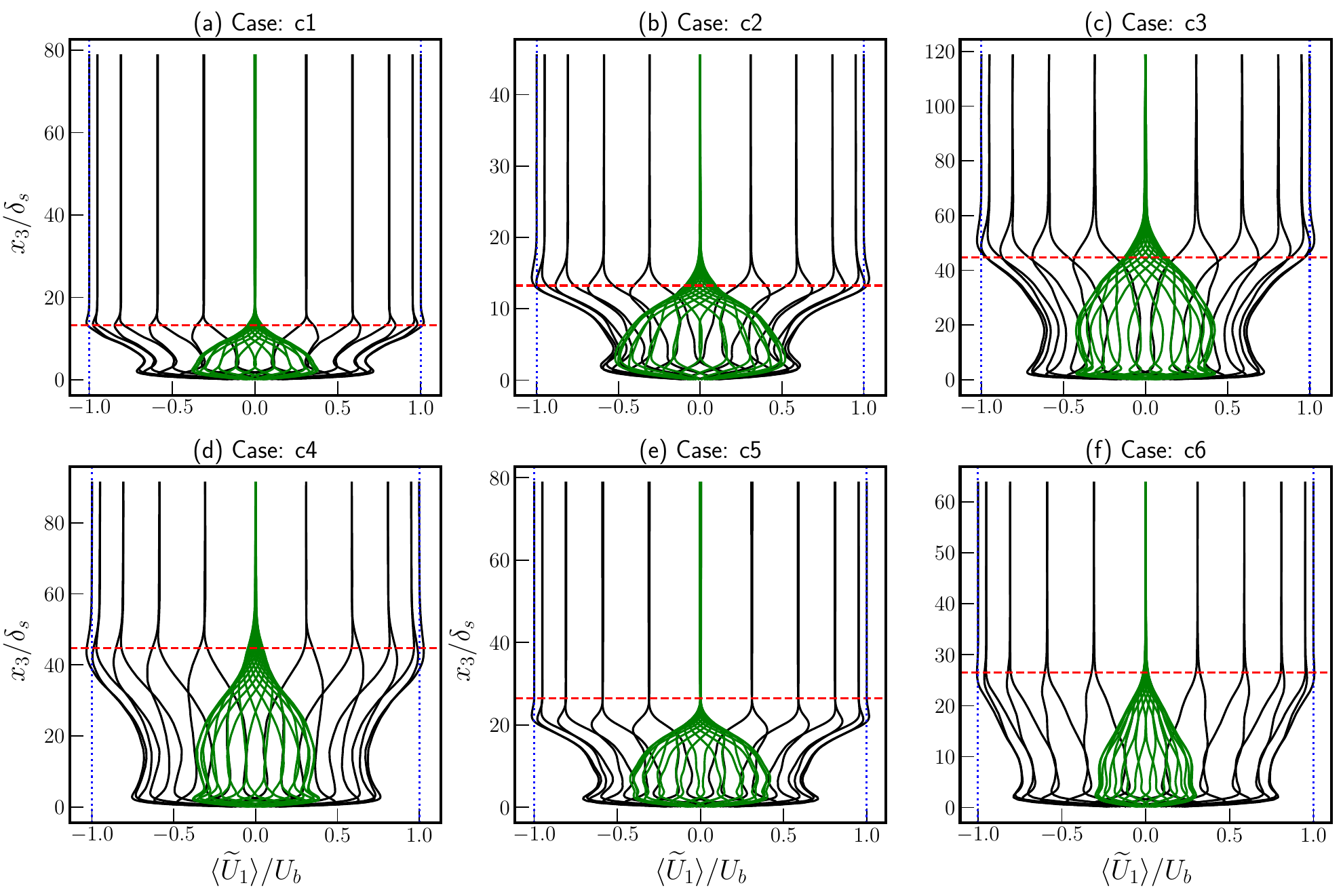}
    \caption{Phase- and planform-averaged velocity profiles for all cases marked using solid-black lines. The velocity deficit profiles are marked using solid-green lines. Vertical blue lines mark the maximum wave orbital velocity for each of the cases, while the horizontal dashed line marks the maximum height of the coral (i.e., $k_s/\delta_s$). Please note the difference in the y-axis extents across the various sub-plots presented in this figure. }
    \label{fig:figure5}
\end{figure}

For a far-field velocity given by $u(\infty,t) = U_b \sin{(\omega t)}$ and no-slip at the wall $u(0,t) = 0$ over a flat wall without any roughness (also famously known as a variation of the Stokes' second problem); the phase-averaged velocity is given by \citet{Stokes1851}

\begin{equation} \label{eq:stokesSolution}
    \widetilde{u}(z,t) = U_b \sin{(\omega t)} + U_b \exp{(-K z)} \sin{(K z - \omega t)},
\end{equation}

\noindent where $K \equiv \sqrt{\omega/(2\nu)}$ and $\widetilde{u}(z,t)$ is the phase- and planform-averaged velocity. It is important to note that the far-field velocity follows a $\sin(\omega t)$ as the driving oscillatory pressure gradient follows a $\cos(\omega t)$ as per equation \ref{eq:nondimNS}. To understand the effect of the coral roughness, the velocity deficit $\widetilde{u}_d (z,t)$ can be defined by

\begin{equation} \label{eq:deficitVelocity}
    \widetilde{u}_d(z,t) = \widetilde{u}(z,t) - \langle \widetilde{U}_1 \rangle (z,t),
\end{equation}

\noindent where the first term on the right-hand side of equation \ref{eq:deficitVelocity} is the Stokes' solution for a flat wall using equation \ref{eq:stokesSolution} and the second term on the right-hand side is the phase- and planform-averaged velocity for a flat wall with corals obtained using the numerical method. The deficit velocity can be understood as the measure of the corals' effect on the phase-averaged velocity. Figure \ref{fig:figure5} shows the numerically obtained velocity profiles for the cases detailed in table \ref{table:table1} using the solid black lines, while solid green lines mark the deficit velocity. By this token, larger deficit components suggest smaller in-canopy flow velocities when the corals are present over the flat wall. Since the planform area for all the cases is similar, most of the differences observed in figure \ref{fig:figure5} can be understood through the varying frontal area detailed in table \ref{table:table2}. Specifically, cases c3 and c6 have relatively large frontal area ratios leading to a smaller in-canopy velocity magnitude compared to the rest of the cases presented in figure \ref{fig:figure5}. While the differences between the in-canopy velocities are small for the other cases, a weak dependence on the morphological parameters can be observed in the wave velocity. Although not shown here, the numerical and analytical solutions detailed in equation \ref{eq:stokesSolution} are in agreement away from the coral roughness as indicated by the vertical dotted blue lines at $U_b = \pm 1.0$. A substantial amount of the velocity profile is observed to be affected by the presence of the coral roughness below the coral canopy, as detailed by the relatively larger deficit velocity components for each of the cases.

\begin{figure}[t]
    \centering
    \includegraphics[width=\linewidth]{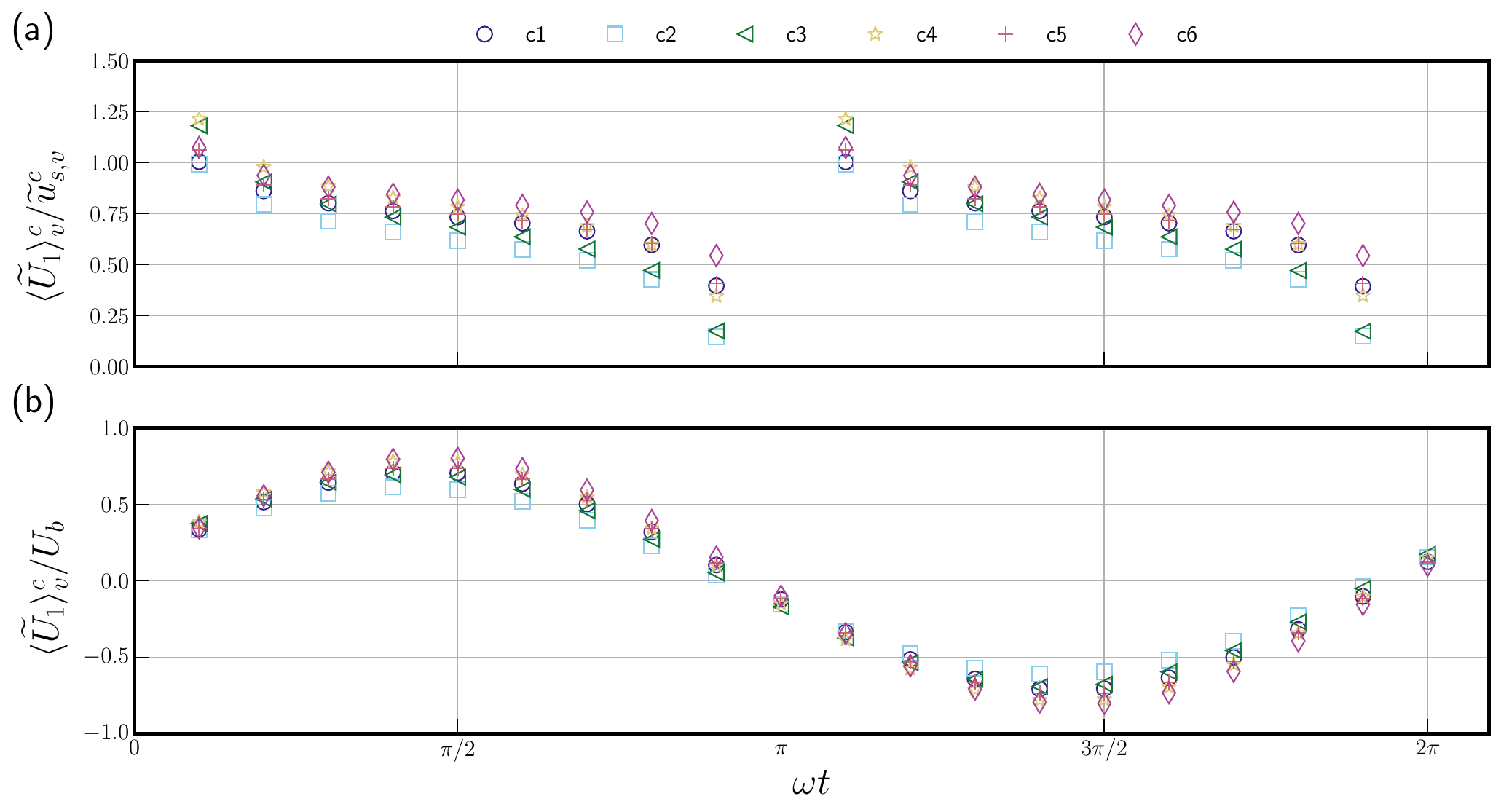}
    \caption{Vertically-integrated, phase- and planform-averaged velocity variations as a function of the wave-phase.}
    \label{fig:figure6}
\end{figure}

To further illustrate the effect of corals on the phase-averaged velocity, the velocity profiles ($\langle \widetilde{U}_1 \rangle$) are vertically integrated over the height of the corals (i.e., horizontal dashed-red line in figure \ref{fig:figure5}) and normalised with the vertically integrated Stokes' solution and maximum wave orbital velocity in figure \ref{fig:figure6}. Case c6 seems to be the least affected by the presence of corals, where the velocity is observed to be consistently higher than that of the other cases. Cases c1, c4, and c5 are observed to undergo similar attenuation of the velocity magnitude during the acceleration and deceleration part of the wave phases, while cases c2 and c3 are affected the most by the presence of corals. The velocity magnitude is substantially attenuated at the peak values, with case c2 showing a reduction of over 40\% while case c6 experiences an attenuation of about 10-15\% when compared to the free-stream wave orbital velocity.

\begin{figure}[H]
    \centering
    \includegraphics[width=\linewidth,trim={0cm 0cm 0cm 0.5cm},clip,scale=0.38]{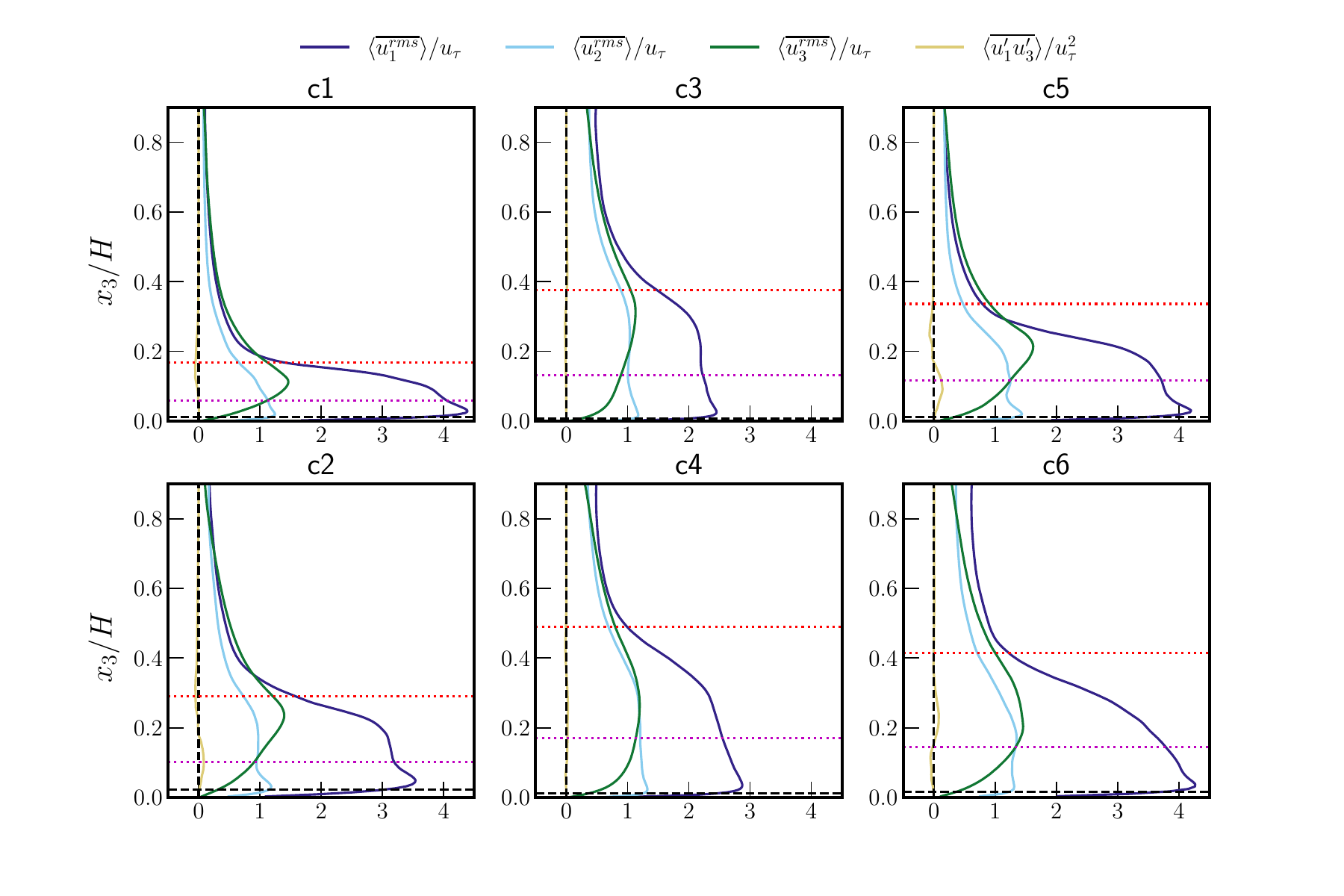}
    \caption{Planform- and time-averaged Reynolds stress and root-mean-squared flow components. The horizontal dashed black line represents the Stokes' boundary layer thickness while the horizontal dotted magenta line represents $0.35 k_s/H$, and the horizontal dotted red line is $k_s/H$, i.e., the peak coral height.}
    \label{fig:figure7}
\end{figure}

While the phase-averaged velocity shows a significant dependence on the coral geometry and the wave Reynolds number, the time-averaged Reynolds stress components as seen in figure \ref{fig:figure7} show a relatively consistent trend across the six cases considered in this study. There is clear dependence on the intensity of the rms velocity profiles on the wave Reynolds number such that cases c1, c2, c5, and c6 exhibit relatively larger rms velocity magnitudes when compared to cases c3 and c4. This difference is found to be relatively consistent within the canopy layer defined as the region between the two horizontal lines in figure \ref{fig:figure7} given by

\begin{equation} \label{eq:canopyLayer}
    \mathcal{L}_c = \beta_s \frac{k_s}{H} \lessapprox x_3 < \frac{k_s}{H},
\end{equation}

\noindent where $\mathcal{L}_c$ is defined as the canopy layer and the fraction $\beta_s = 0.35$ marks the approximate location where the spanwise and vertical rms velocity components are approximately equal in magnitude. Specifically, the spanwise and vertical rms velocity components are observed to be approximately equal to $u_{\tau}$. Despite the difference in the roughness-Reynolds number between cases c1, c2 and c5, c6, no discernible trend was observed for the rms velocity components. Cases c3 and c4 consistently show lower streamwise rms velocity profiles within the canopy region, illustrating the influence of the wave Reynolds number on the flow above the coral roughness compared to the other cases. The spanwise and vertical rms velocity components are 1.0-1.2 times $u_{\tau}$ when there is a streamwise periodic forcing. Below the canopy layer, the vertical rms velocity components develop relatively slowly compared to the spanwise rms velocity component, similar to the trend observed in sheared flows \citep{Pope2000}. For all the cases discussed in this study, there is a consistent peak for the streamwise rms velocity component just above $\delta_s$, beyond which the streamwise rms velocity decreases with most of the peak value showcased inside the canopy region. It must be noted that during the wave phase reversal, the velocities asymptote to zero; consequently, both the numerator and denominator plotted on the y-axis of figure \ref{fig:figure6}(a) become singular. This also explains why the differences start to become large, mainly because of the way the data is presented. Figure \ref{fig:figure6}(b) demonstrates a more discernible trend for the data. These observations are not universal since the synthetic coral geometry is generated using individual corals that stay the same across the cases discussed in table \ref{table:table1}.

Consequently, generalisations should be cautiously made given the relatively small sample size for the coral geometries used in this study. Despite these limitations, the trend in the rms velocity profiles and bulk velocity profiles observed seem to suggest that the heterogeneous composition of the synthetic coral geometries leads to a consistent flow response despite relatively large variations in the morphological characteristics when the appropriate scaling parameters are used. This hypothesis will be tested against other turbulence parameters in the following sections to understand better the similarities and differences observed across the cases presented in table \ref{table:table1}.

The inner-scaled forcing frequency is given by $\omega^+ \equiv  \omega \nu / u_{\tau}^2$, and table \ref{table:table3} lists the forcing frequency for all the cases. As detailed by \citet{Jelly2020}, for a wave forcing frequency that satisfies the condition $\omega^+ > 0.04$, the turbulence in the flow can be assumed to be in a state of `frozen' turbulence. Except for cases c3 and c4, all the cases follow the `frozen' turbulence condition, which suggests that during the wave's peak turbulence generation phase, any generated turbulence will be advected by the flow \citep{Taylor1938}. In this context, peak turbulence generation refers to the production of turbulent kinetic energy through the energy extraction from the mean strain rate via the Reynolds stress terms, which are then transported by various mechanisms \citep{Pope2000}. In the case of periodic forcing over rough walls, turbulence production can occur via two primary mechanisms, namely the mean strain supplying TKE through a classical down-gradient cascade and the work done by the coral roughness on the fluid that generates turbulence at the roughness scale as detailed in \citet{GhodkeApte2016} and \citet{GhodkeApte2018}. This suggests that the viscous time scale and the wave time scale are equally important and that the viscous processes occur over the same time scale as the wave period. Thus, during the wave cycle, any generated turbulence is expected to be advected within the system and dissipated subsequently. Despite the large variation in the wave period for these two cases, the inner-scaled forcing frequencies are similar, and thus, the rms velocity profiles are identical. Specifically, for case c3, close to the wall, there is a small peak at the height of the Stokes boundary layer and within the canopy region. As for case c4, a similar peak at the Stokes' boundary layer height is observed with a secondary peak within the canopy layer. Consequently, in the following subsection, we will discuss in detail the dynamics of in-canopy turbulence.

\begin{table}[t]
    \centering
    \begin{tabular}{c|c}
       Case Name  &  $\omega^+ \equiv \frac{\omega \nu}{u_{\tau}^2}$ \\
       \hline
        c1 & 0.435 \\
        c2 & 0.292 \\
        c3 & 0.011 $ < 0.04$ \\
        c4 & 0.014 $ < 0.04$ \\
        c5 & 0.398 \\
        c6 & 0.430\\ 
    \end{tabular}
    \caption{Inner-scaled wave frequency for the cases simulated in this study.}
    \label{table:table3}
\end{table}

\subsection{Phase averaged turbulence statistics}

\begin{figure}[t]
    \centering
    \includegraphics[width=\linewidth,trim={0cm 0cm 0cm 0.5cm},clip,scale=0.38]{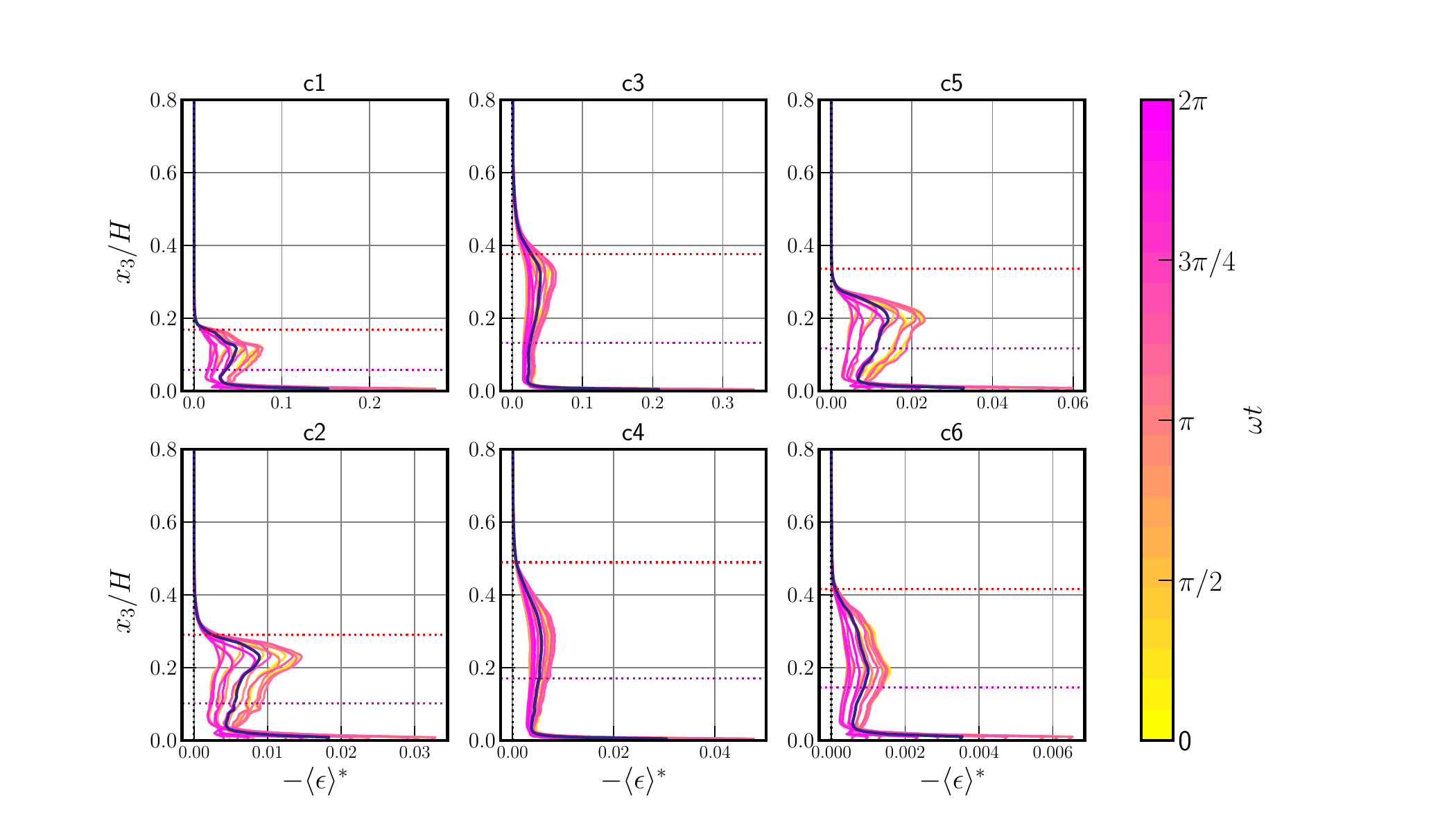}
    \caption{Phase- and planform-averaged TKE dissipation rate (yellow to pink solid lines) as a function of the vertical coordinate and the time-averaged TKE dissipation rate (thicker blue solid line). Here, the non-dimensionalisation is carried out as discussed in \ref{app:appendixA} and denoted by $\langle \epsilon \rangle^*$.}
    \label{fig:figure8}
\end{figure}

\begin{figure}[t]
    \centering
    \includegraphics[width=\linewidth,trim={0cm 0cm 0cm 0.5cm},clip,scale=0.38]{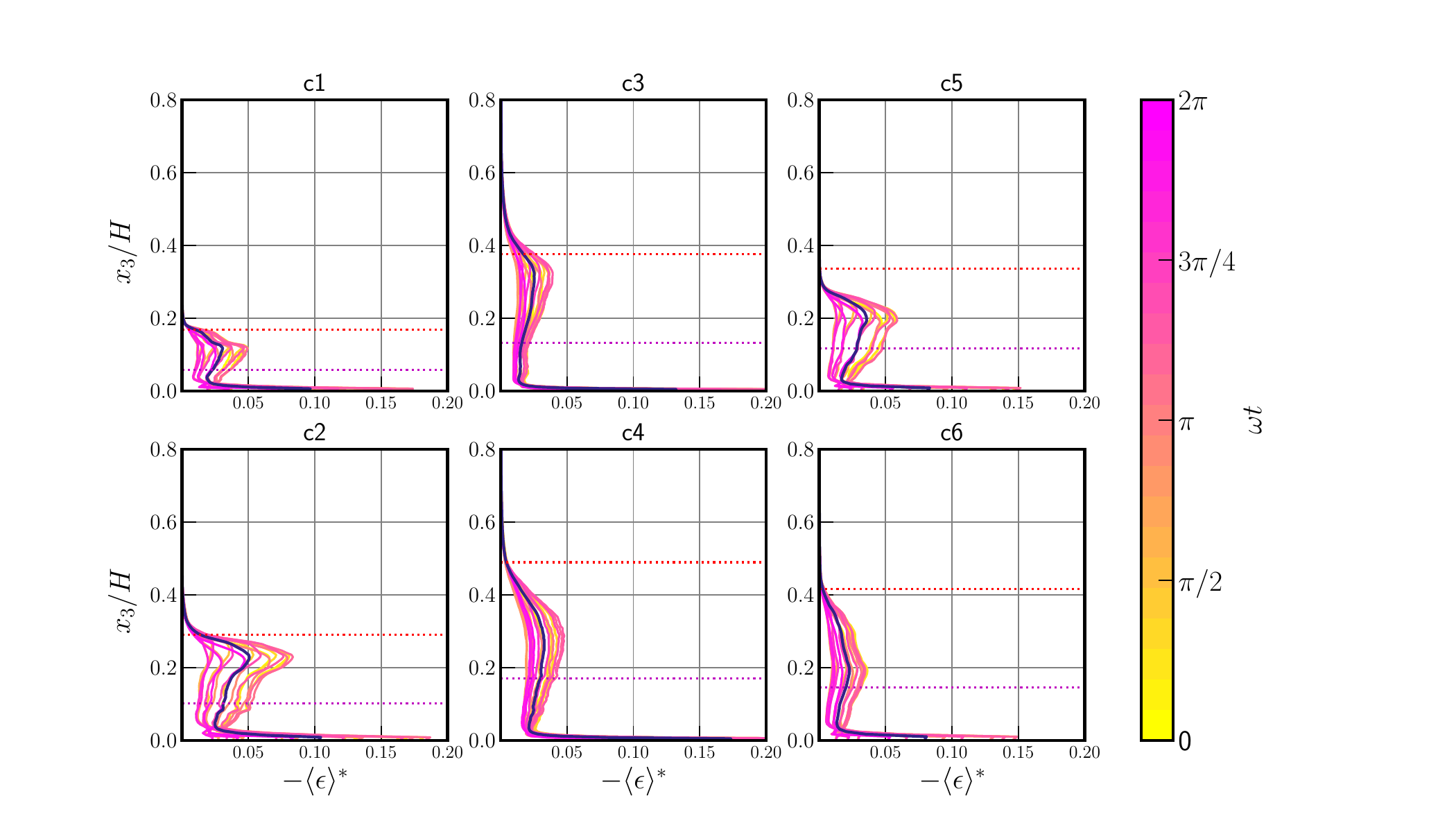}
    \caption{Same as figure \ref{fig:figure8}, inner-scaled TKE dissipation rate where $\langle \epsilon \rangle^* \equiv \langle \epsilon \rangle (\frac{\kappa \nu}{u_{\tau}^4})$.}
    \label{fig:figure9}
\end{figure}

As detailed in \ref{app:appendixA} the TKE dissipation rate ($\langle \epsilon \rangle^*$) is a positive-definite quantity and serves as a sink in the TKE budget and as seen in figure \ref{fig:figure8}, $\langle \epsilon \rangle^*$ scales as expected with $Re_k^b$ and $\Gamma$. Comparing cases c1, c3, c5 against c2, c4, c6 $\langle \epsilon \rangle^*$ is observed to be approximately an order of magnitude different between the two sets of cases. It is important to note that in this representation, $\langle \epsilon \rangle^*$ is scaled using the inertial scaling parameters and not using the inner scaling parameters as would be done conventionally using the kinematic viscosity and the friction velocity. Using the inner scaling for $\langle \epsilon \rangle^*$ as shown in figure \ref{fig:figure9} shows that the differences observed are not so stark between cases c1, c3, c5 and c2, c4, c6, respectively. In fact, inner scaling shows a more consistent trend such that the proportionality with respect to the Reynolds number and relative roughness is relatively more consistent. Despite these differences to have a consistent comparison as detailed in equation \ref{eq:ppTKEBudget}, the outer scaling as shown in figure \ref{fig:figure8} will be used in what follows. This is also motivated by the fact that it is usually easier to estimate the inertial or outer parameters of the flow as opposed to the inner parameters, thus providing a relatively easier method to make a consistent comparison. There is a large peak at the wall, as expected for the oscillatory shear experienced at the bottom wall, and a secondary maximum within the canopy region ($\mathcal{L}_c$), which can be attributed to the coral-induced flow separation. Most of the TKE dissipation is concentrated within the canopy region when the dissipation at the flat wall is ignored for this comparison.
This is an important observation as most of the TKE is generated by virtue of the wake production mechanism \citep{GhodkeApte2018} at a length scale of the coral roughness.
Since the representative length scale varies as a function of the coral density, it is expected that most of the TKE dissipation takes place within this canopy region.
The magnitude of $\langle \epsilon \rangle^*$ for all cases rapidly falls off to zero above $k_s$. The $\langle \epsilon \rangle^*$ is observed to be relatively symmetric when comparing the periodic forcing where each symmetric phase is observed to demonstrate similar values for $\langle \epsilon \rangle^*$ with only minor differences which can be attributed to the heterogeneity in the forward and backward frontal areas experienced by the flow. In the time-averaged sense, a large amount of TKE dissipation is observed as marked by the thick solid blue line for all the cases, which also explains the large deficit velocities observed in figure \ref{fig:figure5}. These results illustrate that even for a synthetically generated coral configuration, most of the energy dissipation occurs within the canopy region, and $\langle \epsilon \rangle^*$ vanishes rapidly above this canopy region unless other generation and/or transport mechanisms are active.

\begin{figure}[t]
    \centering
    \includegraphics[width=\linewidth,trim={0cm 0cm 0cm 0.5cm},clip,scale=0.28]{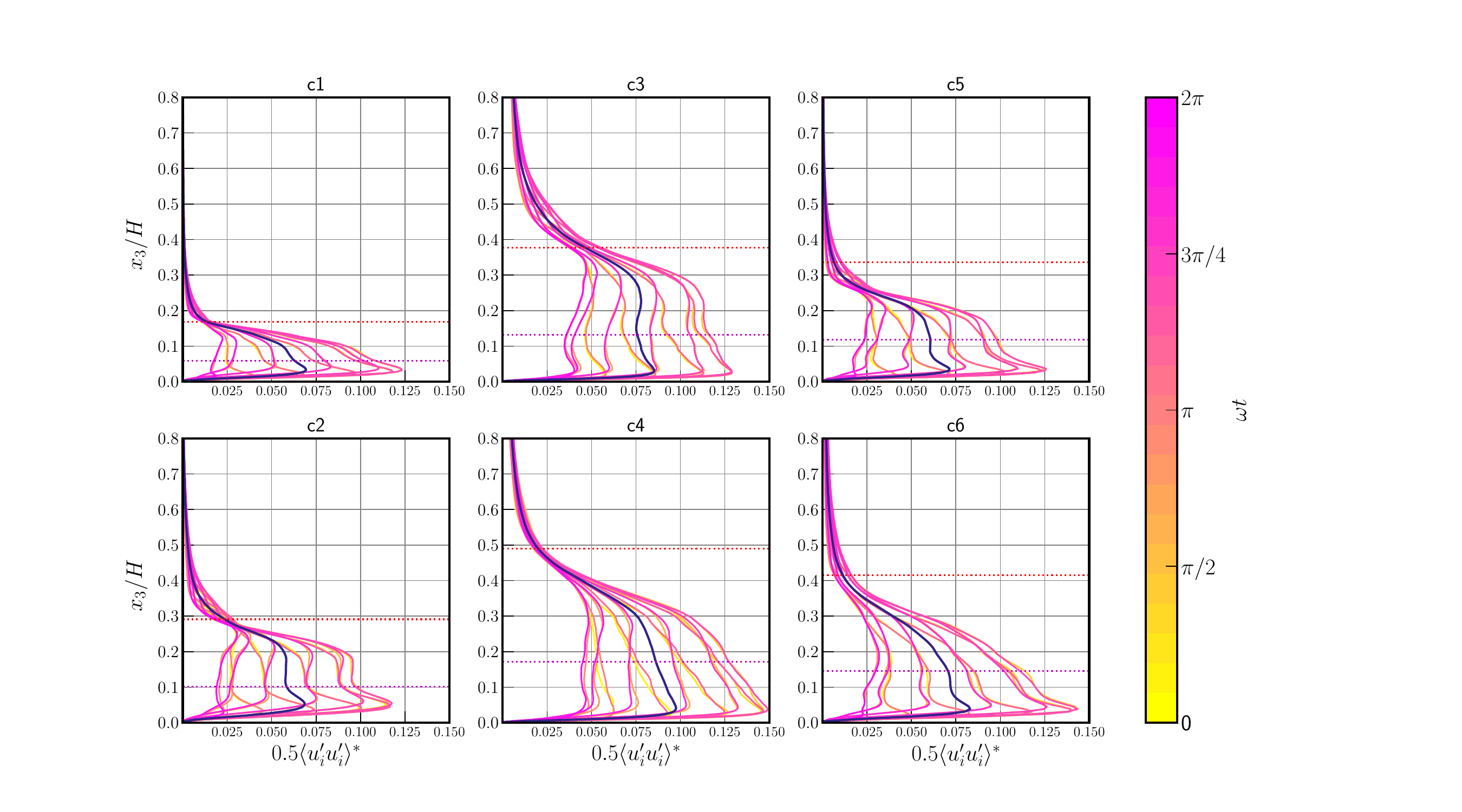}
    \caption{Phase- and planform-averaged TKE (yellow to pink solid lines) as a function of the vertical coordinate and the time-averaged TKE (thicker blue solid line). Here, the TKE is normalised using the square of the maximum wave orbital velocity ($U_b^2$).}
    \label{fig:figure10}
\end{figure}

\begin{figure}[t]
    \centering
    \includegraphics[width=\linewidth,trim={0cm 0cm 0cm 0.5cm},clip,scale=0.28]{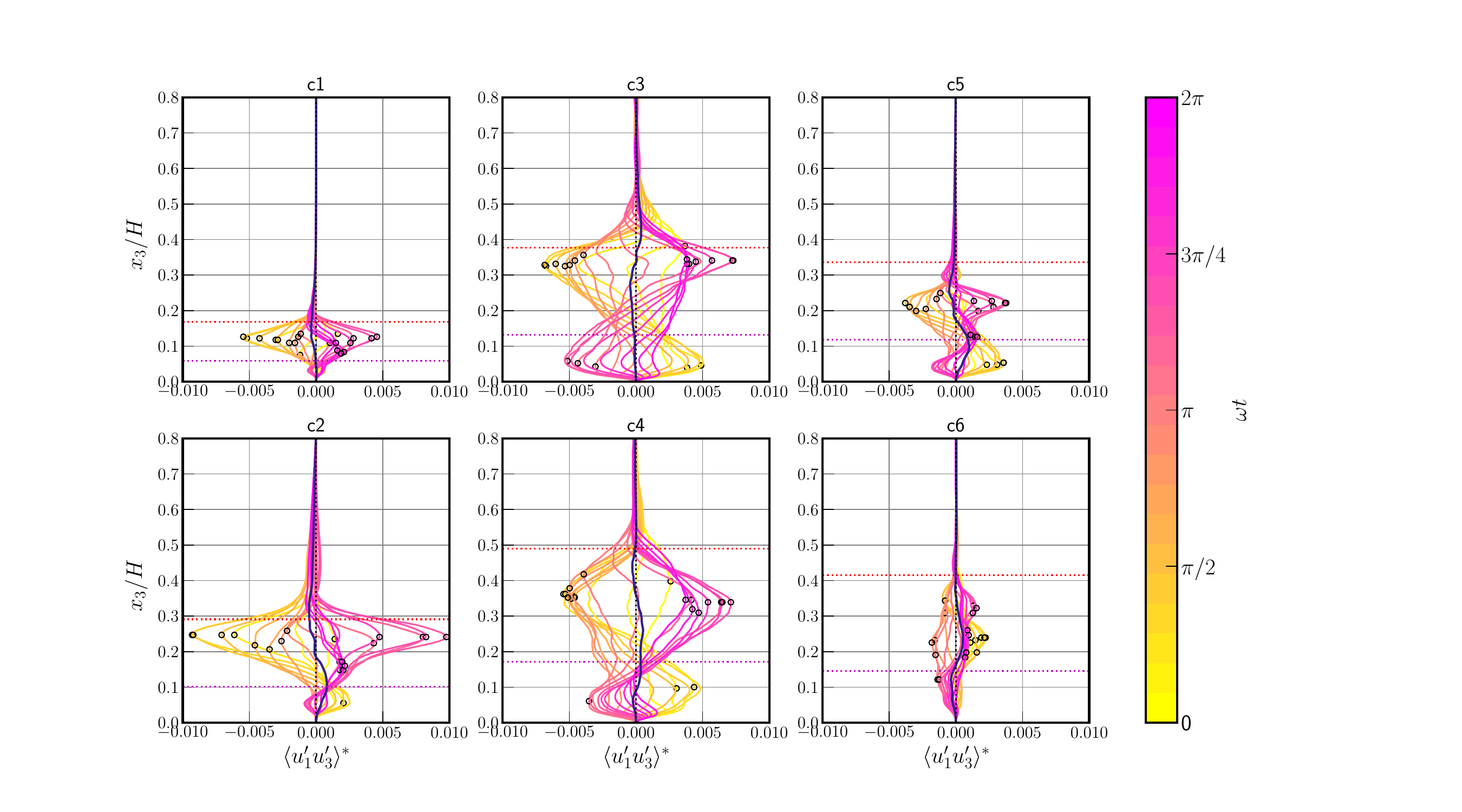}
    \caption{Phase- and planform-averaged vertical Reynolds stress ($ u_1^{\prime} u_3^{\prime}$) as a function of the wave phase. Open black circles mark the maximum value of $\langle u_1^{\prime} u_3^{\prime}\rangle$ for the respective wave phase, and the solid blue line represents the time-averaged value over the entire wave cycle. Here, the stress terms are normalised using the square of the maximum wave orbital velocity ($U_b^2$).}
    \label{fig:figure11}
\end{figure}

The TKE as a function of the wave phase is shown in figure \ref{fig:figure10} using the inertial scaling parameter ($U_b^2$). For all the cases discussed in figure \ref{fig:figure10}, the magnitude of TKE over the wave phase is observed to be wave-direction agnostic, as demonstrated by comparing the magenta solid lines against their yellow counterparts. This is similar to the observations made for the TKE dissipation rate as shown in figures \ref{fig:figure8} and \ref{fig:figure9}. Excluding the TKE peak close to the wall, most of the TKE magnitude is concentrated within the canopy region as defined in equation \ref{eq:canopyLayer}. Across all the cases discussed in this study, the TKE peak is observed to be located at the mid-way point between the canopy height ($\mathcal{L}_c$), while above the top of the coral canopy, the TKE rapidly decreases to a value that is an order of magnitude smaller when compared to the in-canopy TKE magnitude. These observations suggest that for zero mean sinusoidal flows in shallow environments over dense canopies, most of the TKE and the TKE dissipation rate can be observed localised to the central portion of the canopy. This can be further illustrated by the variations in the vertical Reynolds stress ($ u_1^{\prime} u_3^{\prime}$) as detailed in figure \ref{fig:figure11}. For all the cases discussed in this study, the peak vertical Reynolds stress is observed to be located within the canopy layer ($\mathcal{L}_c$) as seen by the clustering of the circle markers in figure \ref{fig:figure11}. For cases c3, c4, and c5, during some wave phases, the peak vertical Reynolds stress is observed closer to the wall and below the canopy region. As for cases c3 and c4, some of the vertical Reynolds stress peak is sustained slightly above the canopy, which can be understood through the large inertial wave Reynolds number compared to the other cases. 

Comparing the peak locations for TKE dissipation rate, TKE, and the vertical Reynolds stress, it is clear that the peaks for all three parameters approximately coincide within the canopy layer. Moreover, the vertical Reynolds stress peaks slightly above the location where the TKE dissipation rate peaks for all the cases, as seen when comparing figures \ref{fig:figure9} and \ref{fig:figure11}. The vertical Reynolds stresses do not exhibit the same wave phase symmetry observed for the TKE dissipation rate and TKE. However, these discrepancies are localised within the vertical direction. For a perfectly symmetric (or sinusoidal) vertical Reynolds stress response, the time-averaged stress integrates to zero for every vertical coordinate. However, as seen in figure \ref{fig:figure11} there is a clear non-zero mean observed for the vertical coordinates until the canopy region. While locally, there is a non-zero mean for the vertical Reynolds stress, when the profile is vertically integrated, it sums out to zero as expected, given that there is no mean flow that balances the shear stress. This suggests that the flow is in equilibrium with the imposed sinusoidal pressure gradient. Above the coral canopy region, all the parameters of interest are an order of magnitude smaller than their in-canopy counterparts, suggesting that the in-canopy aims to set the boundary condition for the flow above it. Close to the flat wall, the vertical Reynolds stress is observed to be in phase with the free stream wave orbital velocity ($U_b$), while within the canopy, peaks are observed to have the exact opposite phase dependence. For all the cases, there is a simultaneous peak observed close to the wall and within the canopy with opposite magnitudes of the vertical Reynolds stresses. As demonstrated by previous studies \citep{Suzukietal2019,JacobsenMcFall2022,Ascencioetal2022,vanRooijenetal2022}, such a dense canopy flow can be modelled effectively using a canopy drag parameterization where the observations made in this study can provide useful building blocks for tuning the model coefficients. While the drag parameterization relies on understanding the bulk effect of the canopy on the flow above it, for unsteady Reynolds Averaged Navier Stokes (URANS) style methods, it can become relevant to model the transport of TKE where the model coefficients can be tuned based on the specific transport mechanisms. Additionally, as detailed by \citet{CondeFriasetal2023} the near bed TKE and vertical Reynolds stress peaks play a significant role in the transport of momentum, TKE, and sediment. When comparing cases c1 and c2 against c3 and c4, the inertial wave Reynolds number is larger by an order of magnitude for the latter cases. As a result, it is expected that with increasing inertial wave Reynolds number, the region closer to the wall experiences larger vertical Reynolds stresses. This is especially consequential when fine sediment is present at the bottom wall which can be then entrained during these parts of the wave cycle. The concentration of the TKE and TKE dissipation within the canopy region also suggests that once suspended, there is a larger potential for such sediment to be dispersed within the canopy region. In coastal systems, surface gravity waves can be accompanied by mean currents, which can then transport both TKE and sediment outside the canopy layer. Overall these findings can provide fruitful insights into the energetics within the canopy region for complex roughness subject to wave motion.

\section{Conclusions}

This study utilised a scalable turbulence-resolving computational framework to model flow around synthetically generated coral reefs over a flat topography. Varied coral canopy parameters were simulated for two inertial wave Reynolds numbers and relative roughness with the aim of better understanding the in-canopy turbulence dynamics. We found that the in-canopy flow is similar for varied coral parameters as long as the free-stream wave orbital velocity is used to normalise the comparison. A weak correlation was observed between the canopy-integrated velocity and the relative roughness; however, given the wide variation in the planform and frontal area for the cases, no discernible conclusions can be made about the overall trend. An important observation in this study suggests that most of the turbulent kinetic energy dissipation occurs within the canopy region, defined as the region between the top of the coral roughness and the location where the spanwise and streamwise root mean squared turbulent velocities are approximately equal to the maximum friction (shear) velocity. Moreover, a similar observation was made for the vertical Reynolds stress, which collectively has broader implications for in-canopy dynamics such as sediment transport, mass transfer rates, and biological activity, to list a few. A weak correlation was also observed between the coral morphology and the turbulent kinetic energy dissipation rate, thus motivating further studies to quantify this connection. In the following future works, a substantial inquiry into first characterising the coral morphology can be envisioned followed by a detailed numerical inquiry into the connections between the morphology and the hydrodynamics response can help bridge the gap in our current understanding of coral reef systems. As a pilot study, this work aimed at adapting the computational framework to study the flow around complex roughness, such as coral reefs, to understand the in-canopy flows. To that end, we find that such a simulation framework can provide further impetus in the development of reduced order models leveraging high-fidelity results and provide deeper insights into the flow around complex roughness in coastal oceanic environments.

\section*{Open Research Section}
The stochastic coral bed generator is freely available through the public repository \url{https://github.com/AkshayPatil1994/turbocor.git}. The stl2sdf generator can be accessed from the public repository \url{https://github.com/AkshayPatil1994/stl2sdf.git}. Some of generated data can be made available upon a reasonable request submitted to the first author (a total of 5 TiB (Terrabytes) of data was generated in this numerical experiment).

\section*{Carbon Footprint Statement}
This work made use of the Snellius supercomputer and had an estimated footprint of 606~kg CO$_2$-equivalent (at least if not higher) using the Green Algorithms (\url{http://calculator.green-algorithms.org/}). The input data used to arrive at these estimates were: Runtime - 1472 hours, Number of cores - 128, Model AMD EPYC 7H12, and Memory available: 256 GiB located in the Netherlands. This is equivalent to taking 1.1 flight(s) from New York (U.S.) to San Francisco (U.S.).

\section*{Acknowledgments}
This work made use of the Dutch national e-infrastructure with the support of the SURF Cooperative using grant no. EINF-6125, supplemented by the Dutch Research Council (NWO). This publication is part of the project "Unraveling the Turbulence Dynamics: Investigating Wave-Induced Turbulence over Corals" (with project number EINF-6125, which is (partly) financed by the Dutch Research Council (NWO).
AP would like to acknowledge the data storage infrastructure provided by the 3D-Geoinformation Research Group, Delft University of Technology. AP would like to thank Ivan Pa\dj en for sharing the wrapwrap application and Pedro Costa for their valuable input in improving the IBM formulation. While preparing this manuscript, the generative AI tool Grammarly was used to check grammar and sentence structures. After using this tool, the authors reviewed and edited the content as needed and take full responsibility for the content of the publication.

%% APPENDICES
\appendix
\section{Derivation of the turbulent kinetic energy budget}
\label{app:appendixA}

Using the velocity decomposition defined in equation \ref{eq:tripleVelocityDecomposition}, we can obtain the wave phase- (henceforth, phase-averaged) and planform-averaged momentum equations to derive the phase- and planform-averaged TKE (ppTKE) equation for this flow. Consider the non-dimensional continuity equation with the triple decomposition given by (dropping the functional dependence parentheses and the $*$ notation when compared to equation \ref{eq:finalNonDimNS})

\begin{equation}
    \label{eq:independentContinuity}
    \partial_i \langle \widetilde{u}_i \rangle + \partial_i \hat{u}_i + \partial_i u_i' = 0,
\end{equation}

\noindent applying the phase- and planform-averaging operators to the equation above results in all the terms vanishing; thus, the continuity equation remains unchanged for this velocity decomposition. It is important to note that the planform-average of the dispersive velocity component is zero by definition (see equation \ref{eq:dispersiveComponent}) and the periodic boundary conditions in the homogeneous directions suggest that any gradients of flow quantities vanish (i.e., $\partial_{1} \langle \cdot \rangle = \partial_{2} \langle \cdot \rangle = 0$). This suggests that, independently, the velocity components are divergence-free for incompressible flows. 

First, we consider the non-dimensional momentum equations decomposed using the phase-averaged and the turbulent velocity components (i.e., equation \ref{eq:velocityDecomposition}) given by

\begin{align}
   & \partial_t \left( \widetilde{u}_i + u_i' \right)  + \Gamma \partial_j \left( \left[ \widetilde{u}_j + u_j' \right] \left[ \widetilde{u}_i + u_i' \right] \right) \notag \\ &= -\Gamma \partial_i \left( \widetilde{p} + p' \right) + \frac{\Gamma}{Re_k^b} \partial_j \partial_j \left( \widetilde{u}_i + u_i' \right) + \cos (t) \delta_{i1},
\end{align}

\noindent where applying the phase-averaging operator after expanding the quadratic non-linear term gives

\begin{equation}
    \label{eq:wavePhaseVelocity}
    \partial_t \widetilde{u}_i + \Gamma \partial_j \widetilde{\tilde{u}_j \tilde{u}_i} = -\Gamma \partial_i \tilde{p} + \frac{\Gamma}{Re_k^b} \partial_j \partial_j \tilde{u}_i - \partial_j \widetilde{u_j' u_i'} + \cos (t) \delta_{i1}, 
\end{equation}

\noindent the equation above is the phase-averaged momentum equation where the energy at the inertial scale is injected by the cosine forcing (i.e., last term on the right-hand side). The turbulent kinetic energy equation can be derived by subtracting equation \ref{eq:wavePhaseVelocity} from the momentum equation in its decomposed form and multiplying the residual equation by $u_i'$ to give

\begin{align}
    & \partial_t u_i' u_i' +  \Gamma \partial_j \tilde{u}_j u_i' u_i' + \Gamma u_i' \partial_j \tilde{u}_j \tilde{u}_i \notag \\ 
    & + \Gamma u_i' u_j' \partial_j \tilde{u}_j + \frac{1}{2}\partial_j u_j' u_i' u_i' + \Gamma u_j' \partial_j \widetilde{\tilde{u}_j \tilde{u}_i} \notag \\ 
    &= -\Gamma \partial_i p' u_i' - \Gamma u_i' \partial_j \widetilde{u_j' u_i'} \notag \\ 
    &+ \frac{\Gamma}{2 Re_k^b} \partial_j \partial_j u_i' u_i' - \frac{\Gamma}{Re_k^b} \partial_j u_i' \partial_j u_i',
\end{align}

\noindent applying the phase-averaging operator to the equation above yields the TKE equation given by

\begin{align}
    \label{eq:tkeBudgetPhaseAvg}
    D_t \tilde{k} & = -\Gamma \widetilde{u_i'u_j'} \partial_j \tilde{u}_i - \frac{\Gamma}{Re_k^b} \widetilde{\partial_j u_i' \partial_j u_i'} + \frac{\Gamma}{Re_k^b} \partial_j \partial_j \tilde{k} \notag \\ & - \partial_j \widetilde{u_j' k} - \Gamma \partial_i \widetilde{p'u_i'}
\end{align}

\noindent where the terms in order are the total rate of change of TKE (i.e., operator defined as $D_t \equiv \partial_t + \Gamma \partial_j \tilde{u}_j$), production of TKE by phase-averaged wave shear, TKE dissipation rate, TKE diffusion by viscosity, turbulent transport of TKE, and pressure-velocity correlations, respectively. It is important to note that as the terms are phase-averaged, they represent the instantaneous TKE balance at a given wave phase. The above derivation uses the identity 

\begin{equation}
    \widetilde{\tilde{f} \tilde{g}} = \tilde{f} \tilde{g} - \overline{\tilde{f} \tilde{g}},
\end{equation}

\noindent where $f$ and $g$ are two flow quantities. The ppTKE equation can now be derived by substituting the velocity decomposition for the phase-averaged component defined in equation \ref{eq:dispersiveComponent} to give

\begin{align}
    \label{eq:ppTKEBudget}
    D_t \langle \tilde{k} \rangle = & - \Gamma \langle \widetilde{u_i' u_j'} \rangle \partial_j \langle \tilde{u}_i \rangle - \frac{\Gamma}{Re_k^b} \langle \widetilde{\partial_j u_i' \partial_j u_i'} \rangle \notag \\ & + \frac{\Gamma}{Re_k^b} \partial_j \partial_j \langle  \tilde{k} \rangle - \partial_j \langle \widetilde{u_j' k} \rangle - \Gamma \partial_i \langle \widetilde{p' u_i'} \rangle \notag \\ &- \Gamma \langle \widehat{u_i' u_j'} \rangle \partial_j \langle \tilde{u}_i \rangle - \frac{\Gamma}{Re_k^b} \langle \widehat{\partial_j u_i' \partial_j u_i'} \rangle \notag \\ & - 2\partial_j \langle \widehat{u_j'k} \rangle - \Gamma \partial_i \langle \widehat{p'u_i'} \rangle
\end{align}

\noindent where the terms in order are the total rate of change of ppTKE, ppTKE production by Reynolds-stress and wave shear, ppTKE rate of dissipation, diffusion of ppTKE by viscosity, turbulent transport of ppTKE, pressure-velocity correlations, production of ppTKE by dispersive stresses and wave shear,  ppTKE rate of dissipation through dispersive stresses, dispersive component of the turbulent transport of ppTKE, and the dispersive component of the pressure-velocity correlations, respectively.

\section{Scaling test for the signed-distance-field generator}
\label{app:appendixB}

\begin{figure}[t]
    \centering
    \includegraphics[width=\linewidth]{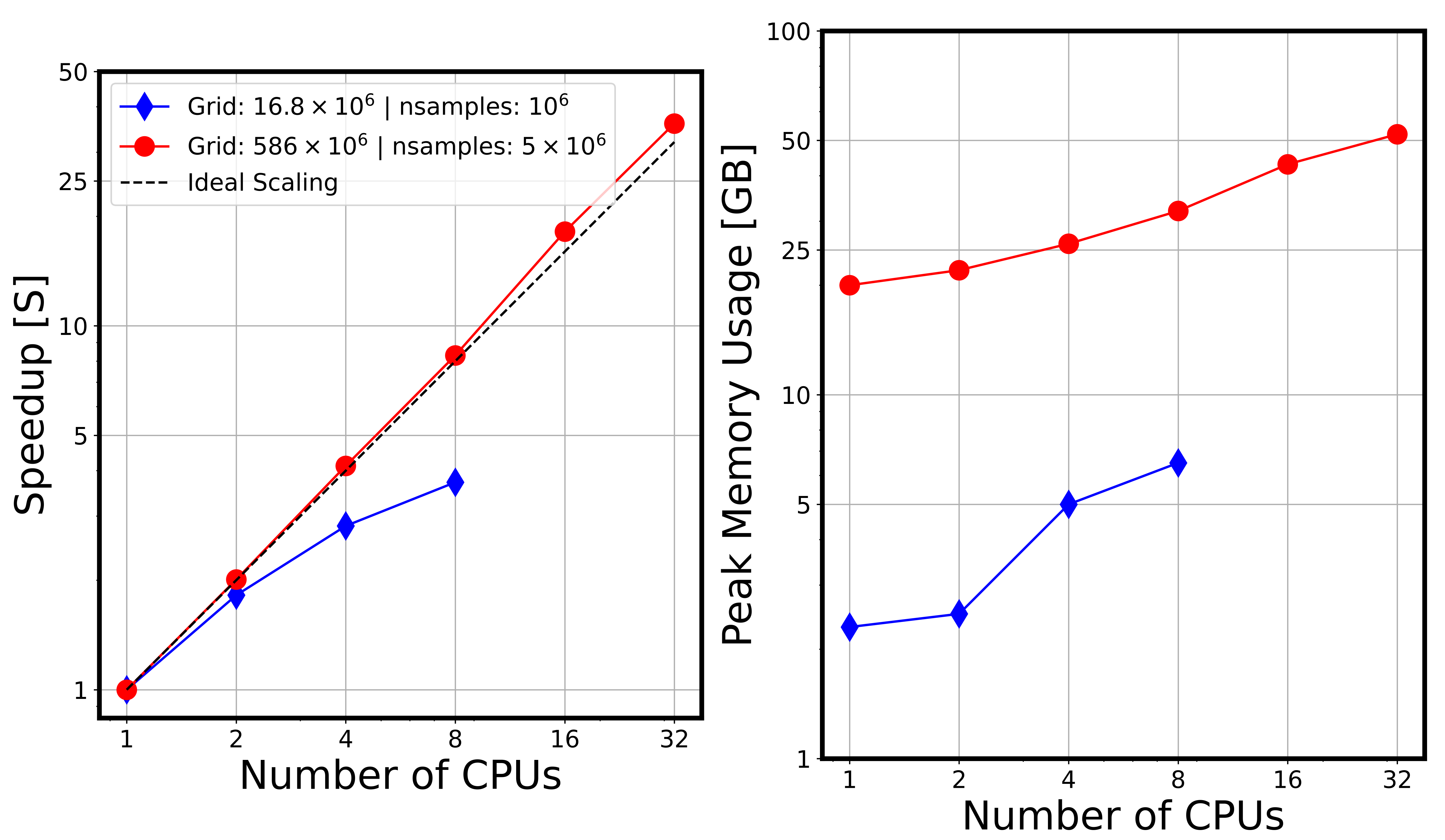}
    \caption{Strong scaling for stl2sdf along with the peak memory usage as detailed in table \ref{table:appendixTable1} along with a smaller scale for comparison.}
    \label{fig:appendixFigure1}
\end{figure}

The signed-distance-field (SDF) generator named stl2sdf was implemented in pure Python and scales up to 2 Billion grid cells over 32 CPUs with a peak memory requirement of approximately 50 GiB for a geometry file of size 57 MiB in total. Table \ref{table:appendixTable1} presents the scaling results for a coral bed where the analysis time represents the time spent on the parallel portion of the code to compute the SDF, and the total time spent represents the total time spent by the code, including the MPI initialisation and file I/O. Figure \ref{fig:appendixFigure1} shows the linear scaling as expected for an SDF algorithm that is easily parallelised as there is no communication between the various domains. The parallelisation uses a simple slab-type decomposition along the $x_1$ flow direction as this is usually the longest dimension in a channel-type flow simulation where stl2sdf is mostly used. The peak memory usage on the right panel indicates sufficiently low memory requirements for a parallel workflow for a limited size of the geometry input. It is important to note that using more cores in stl2sdf can increase the memory requirement non-trivially as the trimesh library \citep{trimesh} used to handle the geometry creates $N$-copies of the geometry when using $N$-processors during the Message-Passing-Interface (MPI) broadcast directive. Despite these minor limitations, the code scales excellently and has been tested for 2 Billion grid points in the computational domain with similar performance and memory requirements for linearly scaled geometry and nsamples.

\begin{table*}[t]
    \centering
    \begin{tabular}{c|c|c|c}
      Number of Processors   & Analysis Time [$s$] & Total Time [$s$] & Peak Memory Usage [GiB] \\
      \hline
        1 & 22584.0 & 22611.8 & 20 \\
        2 & 11244.5 & 1138.2 & 22 \\
        4 & 5476.5 & 5811.7 & 26 \\
        8 & 2723.0 & 3014.5 & 32 \\
        16 & 1243.8 & 1552.9 & 43 \\
        32 & 627.7 & 846.9 & 52 \\
    \end{tabular}
    \caption{Scaling results for generating the SDF for a coral bed with 536.8 Million grid cells with 5 Million sampling points around the geometry.}
    \label{table:appendixTable1}
\end{table*}

%% REFERENCES
\bibliographystyle{elsarticle-harv} 
\bibliography{references}

\end{document}